\documentclass{ws-ijmpa}
\usepackage[super,compress]{cite}
\usepackage{graphicx}
\newcommand{\ETslash}{\ensuremath{/ \hspace{-.7em} E_T}}
\newcommand{\staue}{\tilde \tau_1}
\newcommand{\cha}[1]{\tilde \chi^\pm_{#1}}
\newcommand{\mneu}[1]{\ensuremath{m_{\tilde \chi^0_{#1}}}}

\begin{document}
\markboth{John Ellis}{Prospects for Future Collider Physics}

%
\catchline{}{}{}{}{}
%

\title{Prospects for Future Collider Physics}

\author{John Ellis}

\address{Theoretical Particle Physics and Cosmology Group, Department of Physics, \\
King's College London, Strand, London WC2R 2LS, United Kingdom; \\
Theoretical Physics Department, CERN, CH 1211 Geneva 23, Switzerland\\
John.Ellis@cern.ch}

\maketitle


\begin{abstract}
One item on the agenda of future colliders is certain to be the Higgs boson.
{\it What is it trying to tell us?}
The primary objective of any future collider must surely be to identify
physics beyond the Standard Model, and supersymmetry is one of the most studied options.
{\it Is supersymmetry waiting for us and, if so, can LHC Run 2 find it?}
The big surprise from the initial 13-TeV LHC data has been the appearance of a possible
signal for a new boson $X$ with a mass $\simeq 750$~GeV.
{\it What are the prospects for future colliders if the $X(750)$ exists?}
One of the most intriguing possibilities in electroweak physics would be the discovery
of non-perturbative phenomena. {\it What are the prospects for observing sphalerons
at the LHC or a future collider?}\\
~\\
{\it Contribution to the Hong Kong UST IAS Programme and Conference 
on High-Energy Physics, based largely on personal research with various collaborators.}\\
~\\
KCL-PH-TH-2016-16, LCTS-2016-12, CERN-TH-2016-075

\end{abstract}

\ccode{PACS numbers: 12.15.-y, 12.60.Jv, 14.80.Bn, 14.80.Ly}


\section{The Higgs Boson}	

We already know the mass of the Higgs boson with an accuracy $ \sim 0.2$\%:
\begin{equation}
m_H \; = \; 125.09 \pm 0.21 \pm 0.11 \, {\rm GeV} \, ,
\label{mh}
\end{equation}
where the first (dominant) uncertainty is statistical and the second is systematic~\cite{ATLAS+CMS}.
We can expect that the LHC experiments will reduce the overall uncertainty to
below 100~MeV, setting a hot pace for future collider experiments to follow.
Precise knowledge of the mass of the Higgs boson will be
important for precision tests of the Standard Model - some Higgs decay rates
depend on it quite sensitively - but is also crucial for understanding the stability
of the electroweak vacuum~\cite{EWvac}, as discussed later.

One of the most basic questions about the Higgs boson is whether it is elementary
or composite. In the former case, the large sizes of loop corrections pose the
problem of the naturalness (fine-tuning) of the electroweak scale. The 
solution to this problem that I personally prefer is to postulate an effective cutoff around a TeV due to
supersymmetry, though the absence of supersymmetric particles at the LHC
(so far) is putting a dent in some people's confidence in this solution. The
alternative idea that the Higgs boson is composite has some historical precedents
in its side, namely the composite mesons of QCD and the Cooper pairs of
superconductivity. Early versions of this idea tended to fail electroweak precision
tests and predicted a relatively heavy Higgs-like scalar boson. However, more recent
versions interpret the Higgs boson as a relatively light pseudo-Nambu-Goldstone
boson and interest in composite models may be resurrected if the existence of the $X(750)$
boson is confirmed. For the moment, though, there is as little evidence for composite
models as for supersymmetry.

Under these circumstances, a favoured approach is to assume that the
Higgs boson and all other known particles are described by Standard Model
fields, and parametrise the possible effects of new physics beyond the Standard Model
via higher-dimensional combinations of them, {\it e.g.}, at dimension six~\cite{dim6}:
\begin{equation}
{\cal L}_{eff} \; = \; \sum_n \frac{c_n}{\Lambda^2} {\cal O}_n \, ,
\label{dim6}
\end{equation}
where $\Lambda \gg m_Z, m_W, m_H$ is the mass scale of new physics
and the coefficients $c_n$ help characterise it. They are to be constrained by experiment,
{\it e.g.}, by precision electroweak measurements, Higgs data and measurements of
triple-gauge couplings (TGCs).

\begin{figure}[t!]
\begin{center}
\includegraphics[height=4cm]{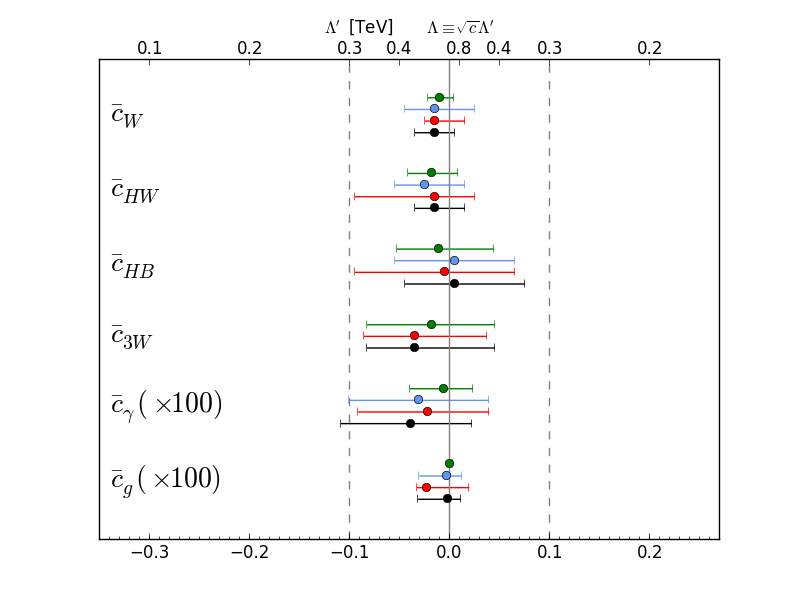}
\includegraphics[height=4cm]{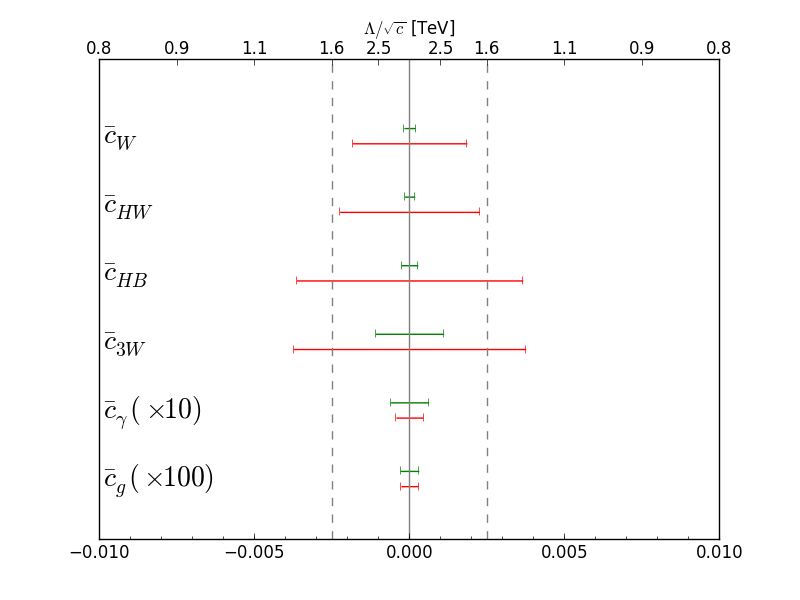} \\
\end{center}   
\caption{\label{fig:dim6}\it 
Left panel: The 95\% CL constraints obtained for single-coefficient fits (green bars),
and the marginalised 95\% ranges for the
LHC signal-strength data combined with the kinematic distributions for associated $H + V$ production
measured by ATLAS and D0 (blue bars), combined with the LHC TGC data (red lines), and the global combination with
both the associated production and TGC data (black bars)~\protect\cite{ESY}. Right panel:
Summary of the 95 \% CL limits on dimension-6 operator coefficients affecting Higgs and TGC observables at FCC-ee~\protect\cite{EY}. 
The individual (marginalised) limits are shown in green (red). 
}
\end{figure}

The left panel of Fig.~\ref{fig:dim6} shows results from one analysis of dimension-six
coefficients~\cite{ESY}, expressed in terms of constraints on the $\Lambda^\prime \equiv \Lambda/\sqrt{c}$
currently provided by these measurements. The lowest (black) error bars are from a global fit
in which all relevant operators are included, whereas the top (green) error bars are from fits
with the operators switched on individually, and intermediate (blue and red) error bars show
the effects of Higgs and TGC measurements, respectively. We see that the
current constraints imply that the $\Lambda^\prime \gtrsim 0.5$~TeV, in general. 

There have been various studies of the sensitivities of future $e^+ e^-$ colliders within this framework.
The right panel of Fig.~\ref{fig:dim6} displays results from one such analysis~\cite{EY}, showing the
prospective sensitivities of measurements at FCC-ee, whose design foresees much greater
luminosities at low energies than the ILC. The upper (green) error bars are for individual operators,
whereas the lower (red) error bars are for a global including all operators. We see that the prospective FCC-ee
constraints would yield sensitivity to $\Lambda^\prime \gg$~TeV, in general.

\begin{figure}[h!]
\begin{center} 
\includegraphics[scale=0.3]{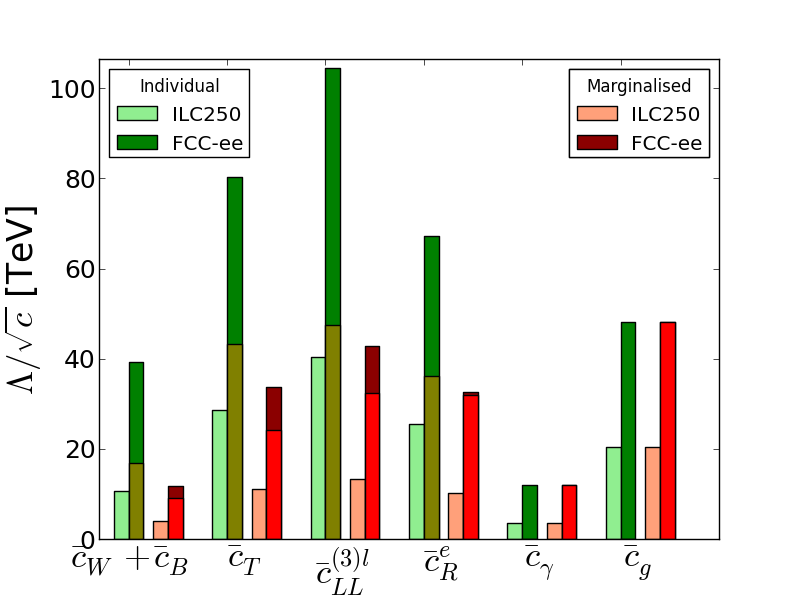}
\includegraphics[scale=0.3]{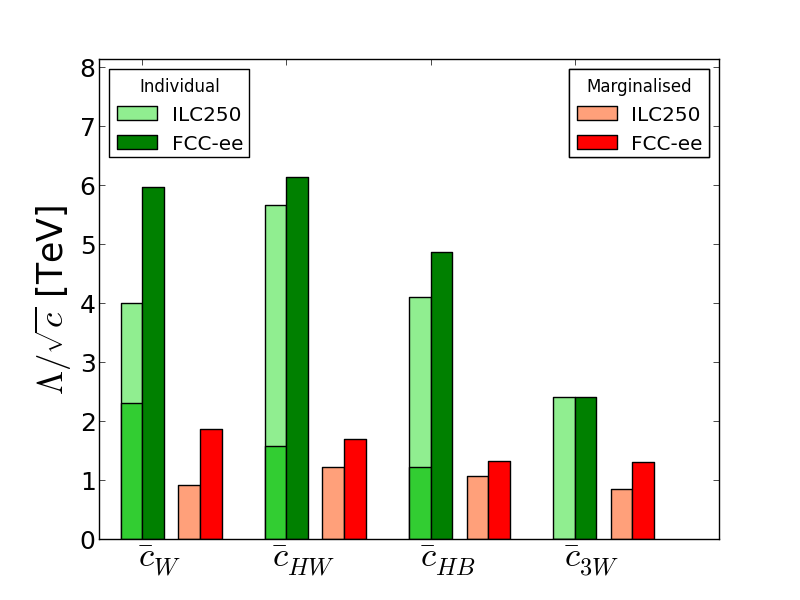}	
\caption{\it Summary of the reaches for the dimension-6 operator coefficients with TeV scale sensitivity, when switched on individually (green) and when marginalised (red), from projected precision measurements at the ILC250 (lighter shades) and FCC-ee (darker shades). The left plot shows the operators that are most strongly constrained by EWPTs and Higgs physics, where the different shades of dark green and dark red represent the effects of EWPT theoretical uncertainties at FCC-ee. The right plot shows constraints
from Higgs physics and TGCs, and the different shades of light green demonstrate the improved sensitivity when TGCs are added at ILC250. Plots from~\protect\cite{EY}.}
\label{fig:ILCFCC-ee}
\end{center}
\end{figure}

Comparisons between the prospective ILC and FCC-ee constraints are shown in Fig.~\ref{fig:ILCFCC-ee},
with the (green) bars on the left representing individual constraints and the right (red) bars marginalised
constraints from a global fit~\cite{EY}. The left panel compares the constraints on a set of dimension-six operators
from Higgs and precision electroweak measurements, with the different darker shadings showing the impact of the
theoretical uncertainties in the latter. We see that FCC-ee has prospective sensitivities in the tens of TeV.
The right panel compares the prospective constraints from Higgs and TGC measurements. We see here 
that FCC-ee could reach into the multi-TeV range, as could the ILC when TGC measurements are included
(lighter shading).

\section{Supersymmetry}

Although the LHC has not yet found any signs of supersymmetric particles, I would argue that
Run~1 of the LHC has actually provided three additional indirect arguments for supersymmetry.
i) In the Standard Model, the measurements of $m_H$ (\ref{mh}) and $m_t$ indicate
 {\it prima facie} that the electroweak vacuum is un/metastable, and supersymmetry would stabilise it.
ii) Simple supersymmetric models predicted successfully the Higgs mass, saying that it should be $< 130$~GeV~\cite{SUSYmH}.
Moreover, iii) simple supersymmetric models predicted successfully that the couplings should be within few \% of
their values in the Standard Model~\cite{EHOW}. These arguments are in addition to the traditional arguments for
supersymmetry based on the naturalness of the electroweak scale, GUTs, string theory, dark matter, {\it etc.}.

Let us review the vacuum stability argument. In the Standard Model the Higgs quartic self-coupling $\lambda$
is renormalised by itself, but the dominant renormalisation is by loops of top quarks, which drive $\lambda < 0$
at some scale $\Lambda$~\cite{EWvac}:
\begin{equation}
\log \frac{\Lambda}{\rm GeV} \; =\; 11.3 + 1.0 \left( \frac{M_h}{\rm GeV} - 125.66 \right) - 1.2 \left( \frac{m_t}{\rm GeV} - 173.10 \right)
+ 0.4 \left( \frac{\alpha_3 (M_Z) - 0.1184}{0.0007} \right) \, .
\label{Lambda}
\end{equation}
The current experimental values of the Higgs mass (\ref{mh}), the official world average top quark mass 
$m_t = 173.34 \pm 0.27 \pm 0.71$~GeV~\cite{WAmt} and the QCD coupling 
$\alpha_3 (M_Z) = 0.1177 \pm 0.0013$~\cite{WAalphas} indicate that the Higgs self-coupling
$\lambda$ turns negative at $\ln (\Lambda/{\rm GeV}) = 10.0 \pm 1.0$ within the Standard Model.
This turndown implies that our present electroweak vacuum is in principle unstable, though its
lifetime may be much larger than the age of the Universe.
However, even in this case there is a problem, since most of the initially hot Universe would not
have cooled down into our electroweak vacuum~\cite{hotU}.

This problem would be completely avoided in a supersymmetric extension of the Standard Model,
where the effective potential is guaranteed to be positive semidefinite. Indeed, one can argue that
vacuum stability may require something very like supersymmetry~\cite{ER}.
Unfortunately, there are many possible supersymmetric extensions of the Standard Model and
no signs in superspace, and we do not know which superdirection Nature may have taken.

What do the data tell us? In the absence of any clues, we use the available electroweak, flavour,
Higgs, LHC and cosmological dark matter constraints in global fits to constrain the parameters of specific 
supersymmetric models.

The simplest possibility is to consider models with universal soft supersymmetry breaking at
some input GUT scale. The scenario in which universality is assumed for the soft supersymmetry-breaking
gaugino masses and those of all the
scalar partners of Standard Model particles and the Higgs multiplets is called the constrained 
minimal supersymmetric extension of the Standard Model (CMSSM)~\cite{CMSSM}, and models in which this 
assumption is relaxed for the Higgs multiplets are called non-universal Higgs models (NUHM1,2)~\cite{NUHM}.
These models are under quite strong pressure from the LHC, with $p$ values $\sim 0.1$~\cite{MCCMSSM, MCNUHM2}.
On the other hand, a model in which the soft supersymmetry-breaking masses are treated
as phenomenological inputs at the electroweak scale (the pMSSM) is less strongly constrained by LHC
data. For example, assuming limited universality motivated by upper limits on flavour-changing neutral interactions
(the pMSSM10), one finds a higher value of $p \sim 0.3$~\cite{MCpMSSM}.

\begin{figure}[t!]
\begin{center}
\includegraphics[height=5cm]{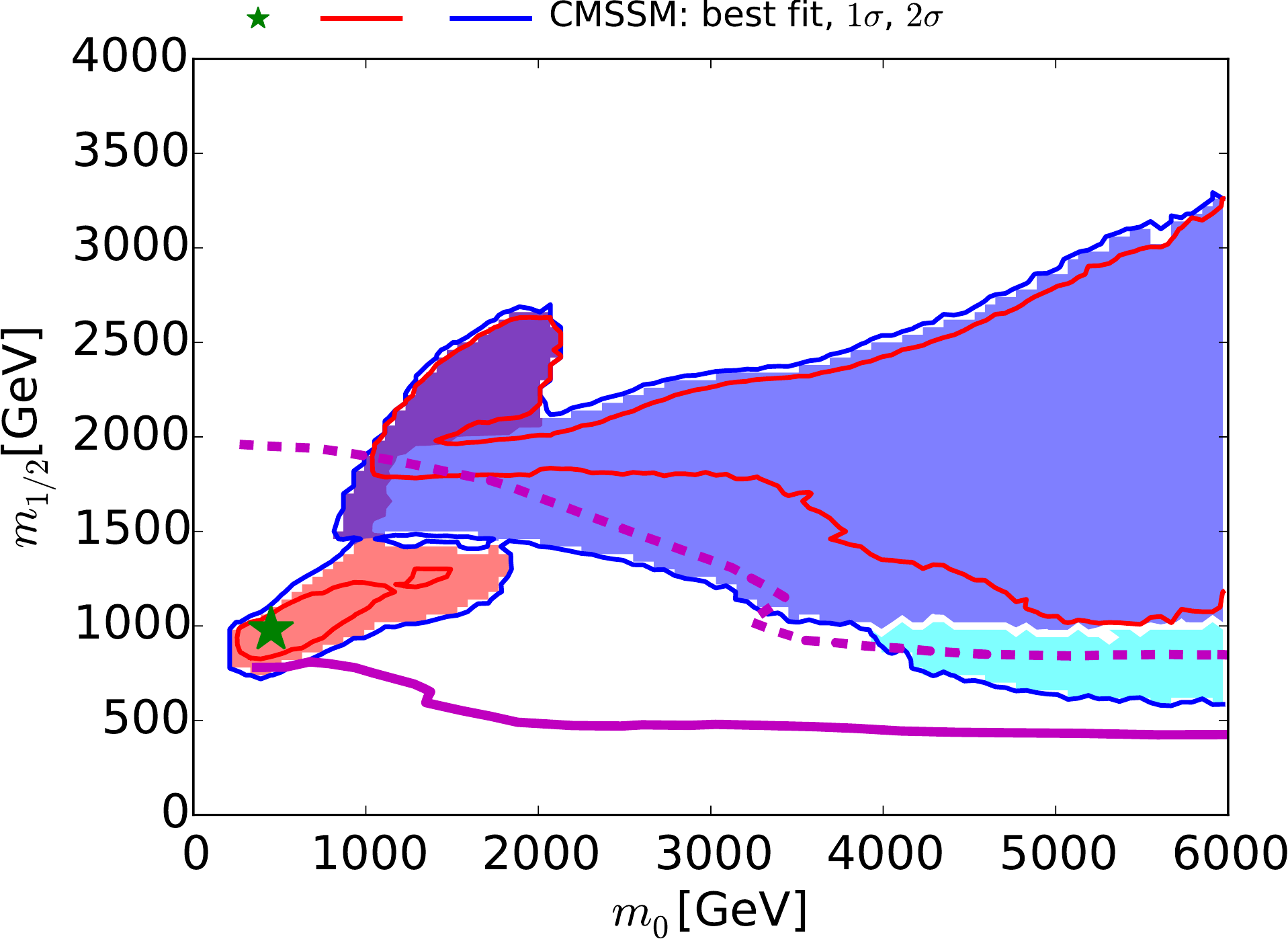}
\end{center}   
\caption{\label{fig:CMSSM}\it 
The $(m_0, m_{1/2})$ planes in the CMSSM. Regions in which different mechanisms bring the dark matter
density into the allowed range are shaded as described in the text~\protect\cite{MCDM}. The red and blue
contours represent the 68 and 95\% CL
contours, with the green star indicating the best-fit point. The solid purple contour
shows the current LHC 95\% exclusion from $\ETslash$ searches, and
the dashed purple contour shows the prospective 5-$\sigma$
discovery reach for ~$\ETslash$ searches at the LHC with 3000/fb at 14~TeV, 
which corresponds approximately to the 95\% CL exclusion
sensitivity with 300/fb at 14~TeV.
}
\end{figure}

Specifically, assuming that the cosmological dark matter is provided by the lightest neutralino,
in the CMSSM the dark matter density constraint provides an upper limit on the
supposedly universal fermion mass $m_{1/2}$ for fixed scalar mass $m_0$, whereas at low values of $m_0$
and $m_{1/2}$ there is tension between the LHC searches for missing-energy events and the
anomalous magnetic moment of the muon, $g_\mu -2$. The $\sim 3 \sigma$ discrepancy between the experimental
measurement and the Standard Model calculation could be resolved via low-mass supersymmetry, but this
cannot be achieved within the CMSSM and related models. Fig.~\ref{fig:CMSSM} displays the $(m_0, m_{1/2})$
plane in the CMSSM, with the region favoured at the 68\% CL bounded by the red contour and that allowed
at the 95\% CL bounded by the blue contour, and the best-fit point indicated by a green star. 
The region in which coannihilation with the lighter ${\tilde \tau}$
slepton brings the dark matter density into the range allowed by cosmology is shaded pink, that where rapid
annihilation via direct-channel heavy Higgs bosons is shaded dark blue, that where both mechanisms are
important is shaded purple, and the stop coannihilation region
is shaded lighter blue~\cite{MCDM}. We see that the current LHC constraint is important at low $(m_0, m_{1/2})$,
and that the estimated future LHC sensitivity covers all the ${\tilde \tau}$ coannihilation region and part of the
rapid heavy Higgs annihilation and stop coannihilation regions.

\begin{figure}[ht!]
\centerline{
\includegraphics[height=5cm]{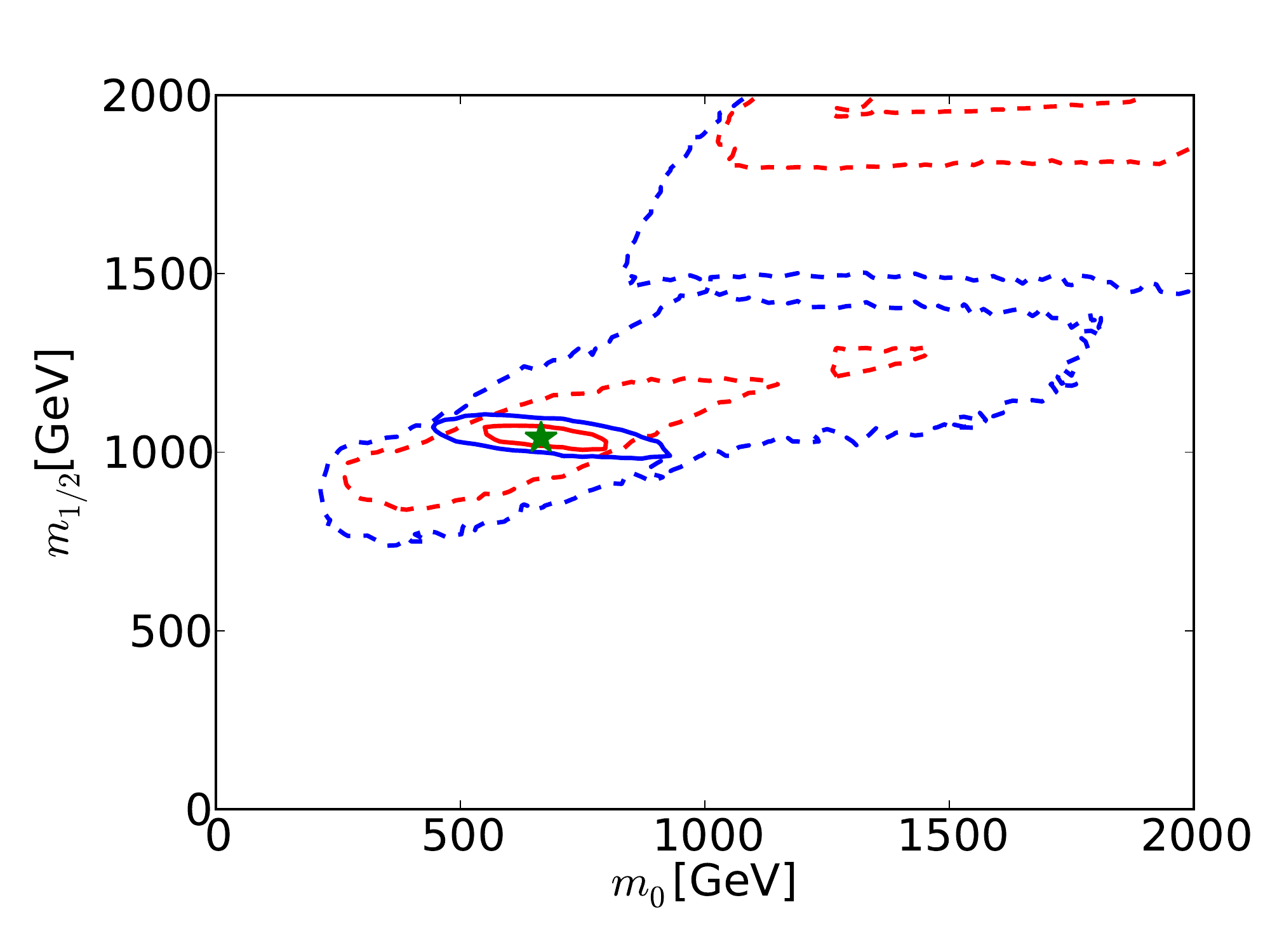}
}
\caption{\label{fig:interplay}\it
The 68 and 95\% CL regions in the $(m_0, m_{1/2})$ planes (solid red and blue lines)
obtained by combining prospective cross-section, $\ETslash$ and  jet measurements with 3000/fb 
of luminosity at the LHC at a centre-of-mass energy of 14~TeV with the current global fit (here shown as dashed lines).
Plot from~\protect\cite{Interplay}.}
\end{figure}

Let us be optimistic, and assume that Nature is described by the current best-fit point in the CMSSM, namely the
green star inside the stau coannihilation region in Fig.~\ref{fig:CMSSM}. In this case it would be possible not
only to discover supersymmetry in future runs of the LHC, but also to measure some of its parameters quite
accurately, as seen in Fig.~\ref{fig:interplay}~\cite{Interplay}. The prediction of the supersymmetric mass scale and such a
detailed confrontation between direct and indirect constraints on supersymmetry would provide tests of
the underlying theory akin to those of the Standard Model provided by direct and indirect constraints on the
masses of the top quark and the Higgs boson.

\begin{figure}[t!]
\begin{center}
\includegraphics[height=4cm]{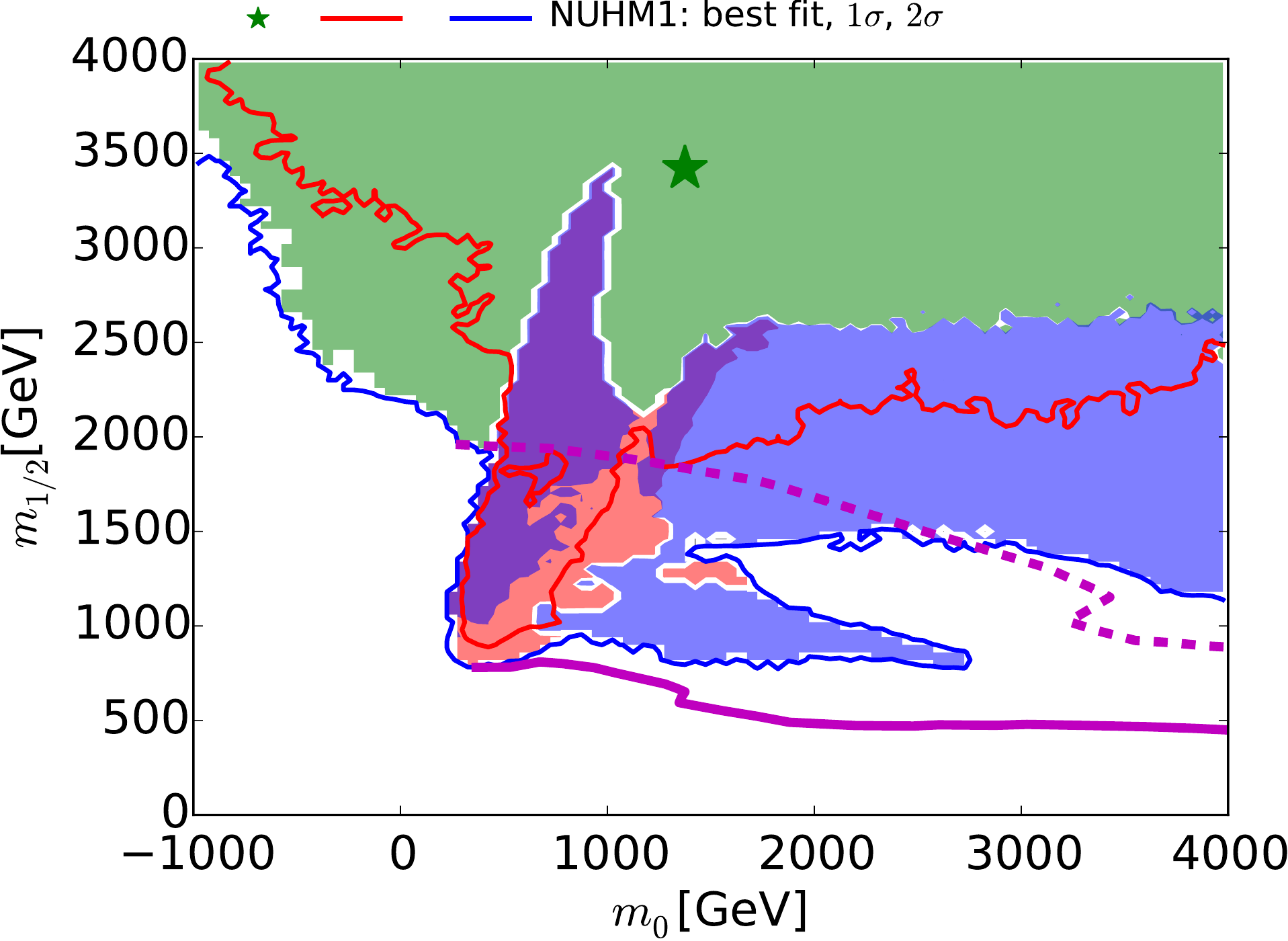}
\includegraphics[height=4cm]{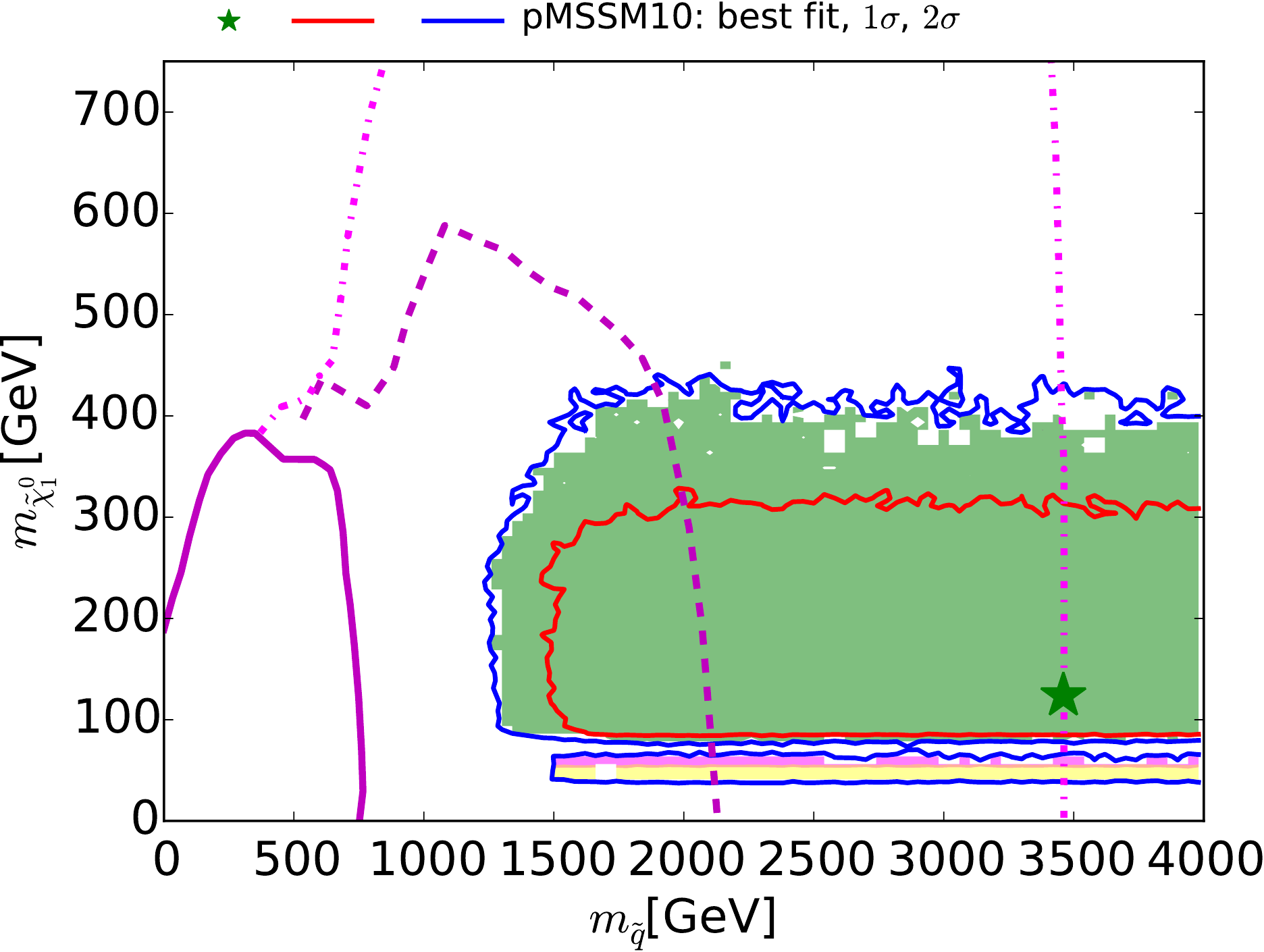} \\
\end{center}   
\caption{\label{fig:otherplanes}\it 
Left panel: The $(m_0, m_{1/2})$ plane in the NUHM1.
Right panel: The $(m_{\tilde q}, \mneu1)$ plane in the pMSSM10.
In the green regions the dark matter density is brought into the allowed range by
chargino coannihilation and in the pink and yellow strips in the right panel by rapid annihilation via the $h$ and $Z$ poles:
the other colours in the left panel have the same significances as in Fig.~\protect\ref{fig:CMSSM}. Plots from~\protect\cite{MCDM}.}
\end{figure}

\begin{figure*}[htb!]
\begin{center}
\resizebox{7.5cm}{!}{\includegraphics{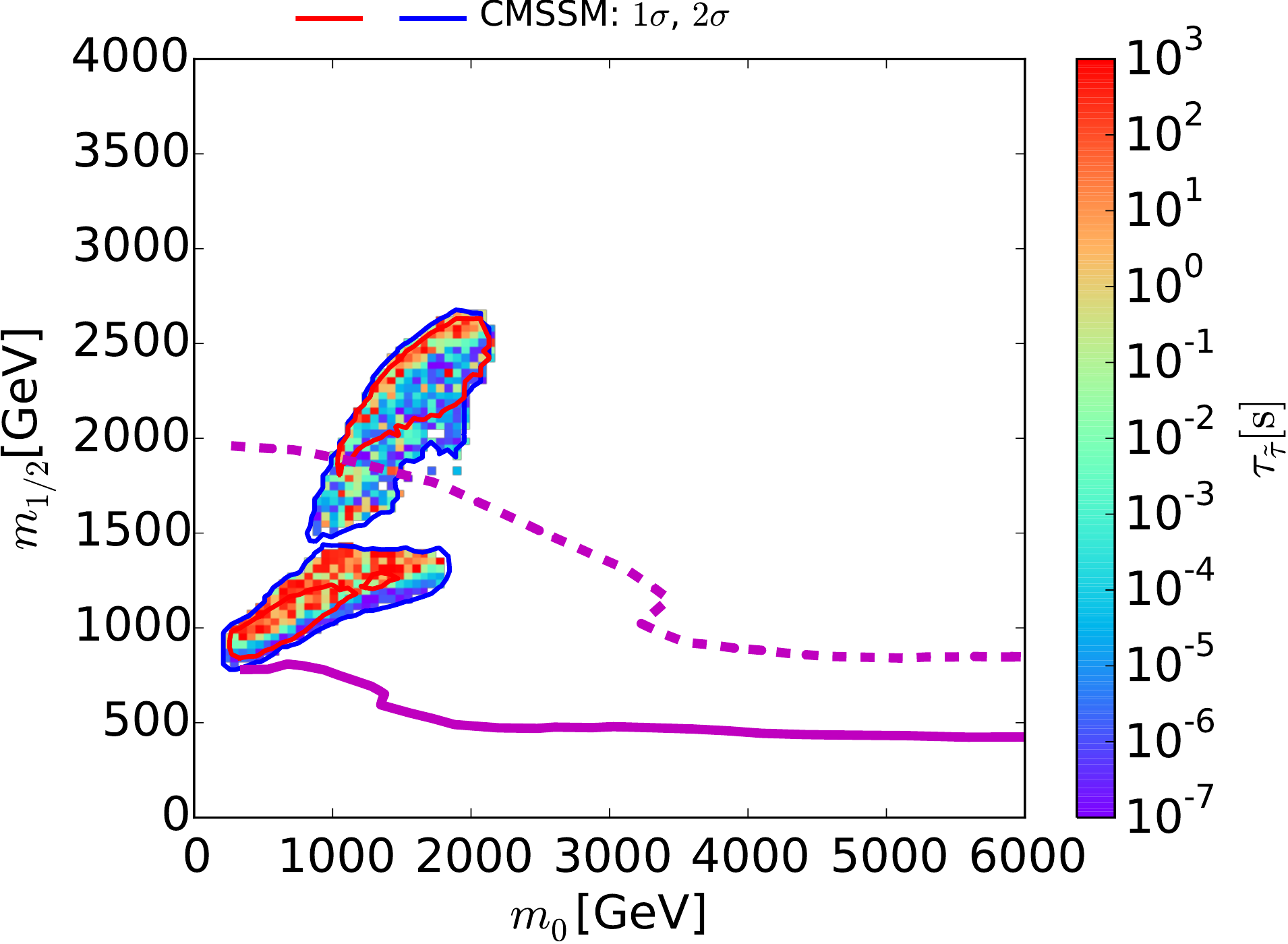}}
\end{center}
\vspace{-0.5cm}
\caption{\it The $(m_0, m_{1/2})$ plane in the CMSSM,
showing the regions where the lowest-$\chi^2$ points within the 95\% CL
region that have $10^3$s $> \tau_{\staue} > 10^{-7}$s: 
the lifetimes of the $\staue$ at these points are colour-coded, as indicated in the legends~\protect\cite{MCDM}.
Also shown in these panels as solid purple contours are the current LHC 95\% exclusions from $\ETslash$ searches
in the ${\staue}$ coannihilation regions,
and as dashed purple contours the prospective 5-$\sigma$
discovery reaches for $\ETslash$ searches at the LHC with 3000/fb at 14~TeV, corresponding approximately to the 95\% CL exclusion
sensitivity with 300/fb at 14~TeV. The sensitivities of LHC searches for metastable
${\staue}$'s in the ${\staue}$ coannihilation region are expected to be
similar~\protect\cite{nigh}.}
\label{fig:lifetime}
\end{figure*}

The left plane of Fig.~\ref{fig:otherplanes} shows the $(m_0, m_{1/2})$ plane in the NUHM1, using the same
colouring scheme as in Fig.~\ref{fig:CMSSM}. We see again that the LHC should be able to explore an
interesting area of the NUHM1 parameter space~\cite{MCCMSSM}, 
and the same is true of the NUHM2 parameter space (not shown)~\cite{MCNUHM2}. 
What would be a key distinctive signature of
supersymmetry in the CMSSM and the NUHM1,2? Much of the parameter spaces of these models
accessible to the LHC lies in the stau coannihilation region, where the mass difference between the lighter
stau ${\tilde \tau_1}$ and the lightest neutralino ${\tilde \chi_1^0}$ is quite small. 
In such a case, the lifetime of the next-to-lightest supersymmetric
particle (NLSP), the ${\tilde \tau_1}$, may be quite long, as seen in Fig.~\ref{fig:lifetime}~\cite{MCDM}. Thus, possible signatures
could include long-live charged particles that decay outside the detector or with a separated decay vertex within it~\cite{nigh}.

\begin{figure}[t!]
\begin{center}
\includegraphics[height=5cm]{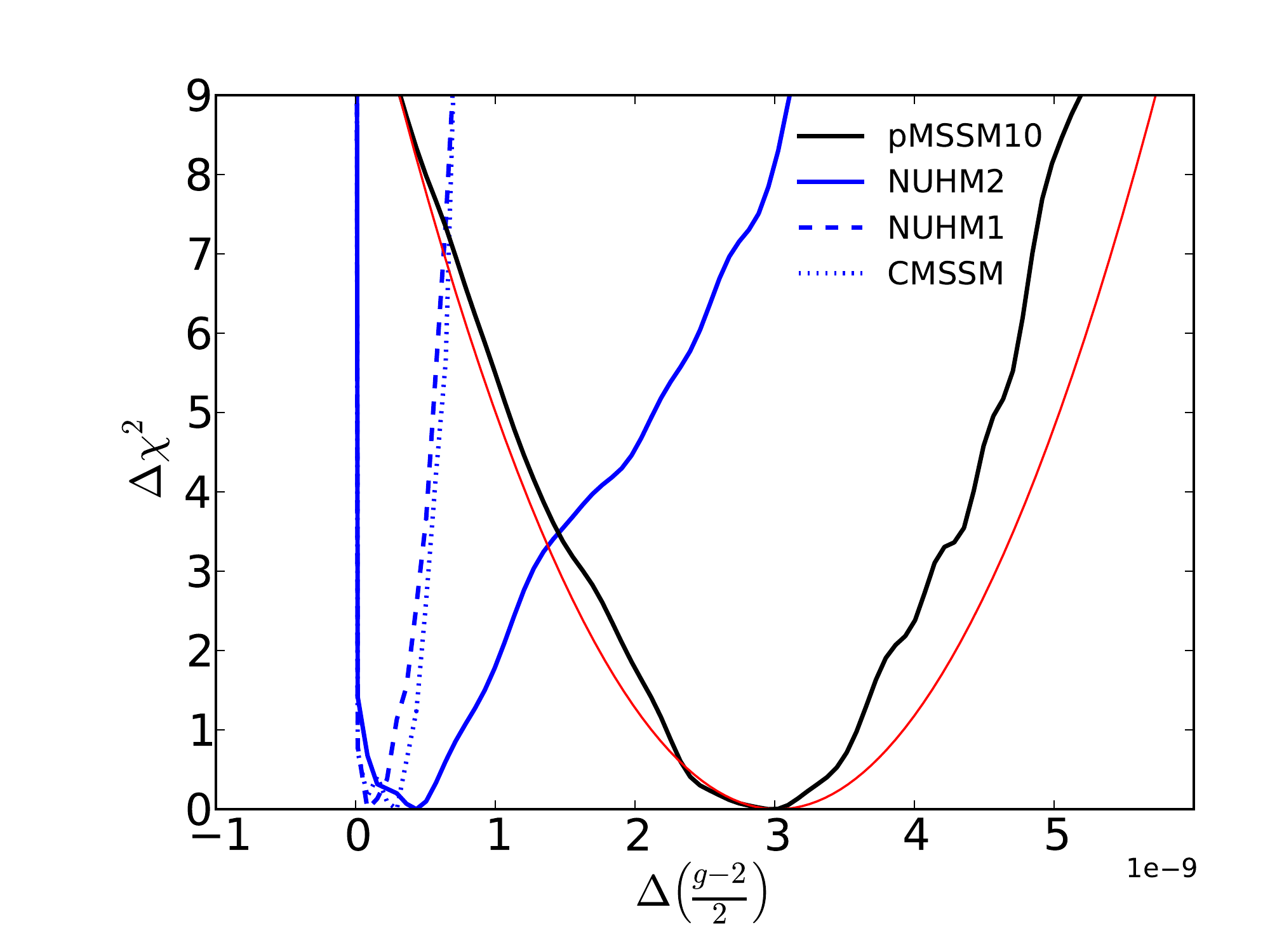}
\end{center}   
\caption{\label{fig:g-2}\it 
The $\chi^2$ likelihood functions for the anomalous magnetic moment of the muon, $g_\mu - 2$,
in the CMSSM, NUHM1, NUHM2 and pMSSM10, taking account of LHC Run~1 and other constraints,
as described in~\protect\cite{MCpMSSM}.
}
\end{figure}

The situation is rather different within the pMSSM10, whose $(m_{\tilde q}, m_{\chi^0_1})$ plane is shown
in the right panel of Fig.~\ref{fig:otherplanes}~\cite{MCDM}. We see again that future runs of the LHC have a fair chance
of discovering supersymmetry also in this scenario (the dashed line is for $m_{\tilde q} \ll m_{\tilde g}$
and the dash-dotted line for $m_{\tilde g} = 4.5$~TeV), but we do not expect a long-lived charged particle signature.
However, the pMSSM10 can resolve the tension between LHC searches and the measurement of $g_\mu - 2$~\cite{MCpMSSM}.
Fig.~\ref{fig:g-2} shows that, whereas the CMSSM and related models (blue curves) predict values of $g_\mu -2$
that are very similar to those in the Standard Model, the pMSSM10 (black curve)~\cite{MCpMSSM} can accommodate the
experimental value (red curve) without falling foul of the LHC constraints.

\begin{figure}[t!]
\begin{center}
\includegraphics[height=4cm]{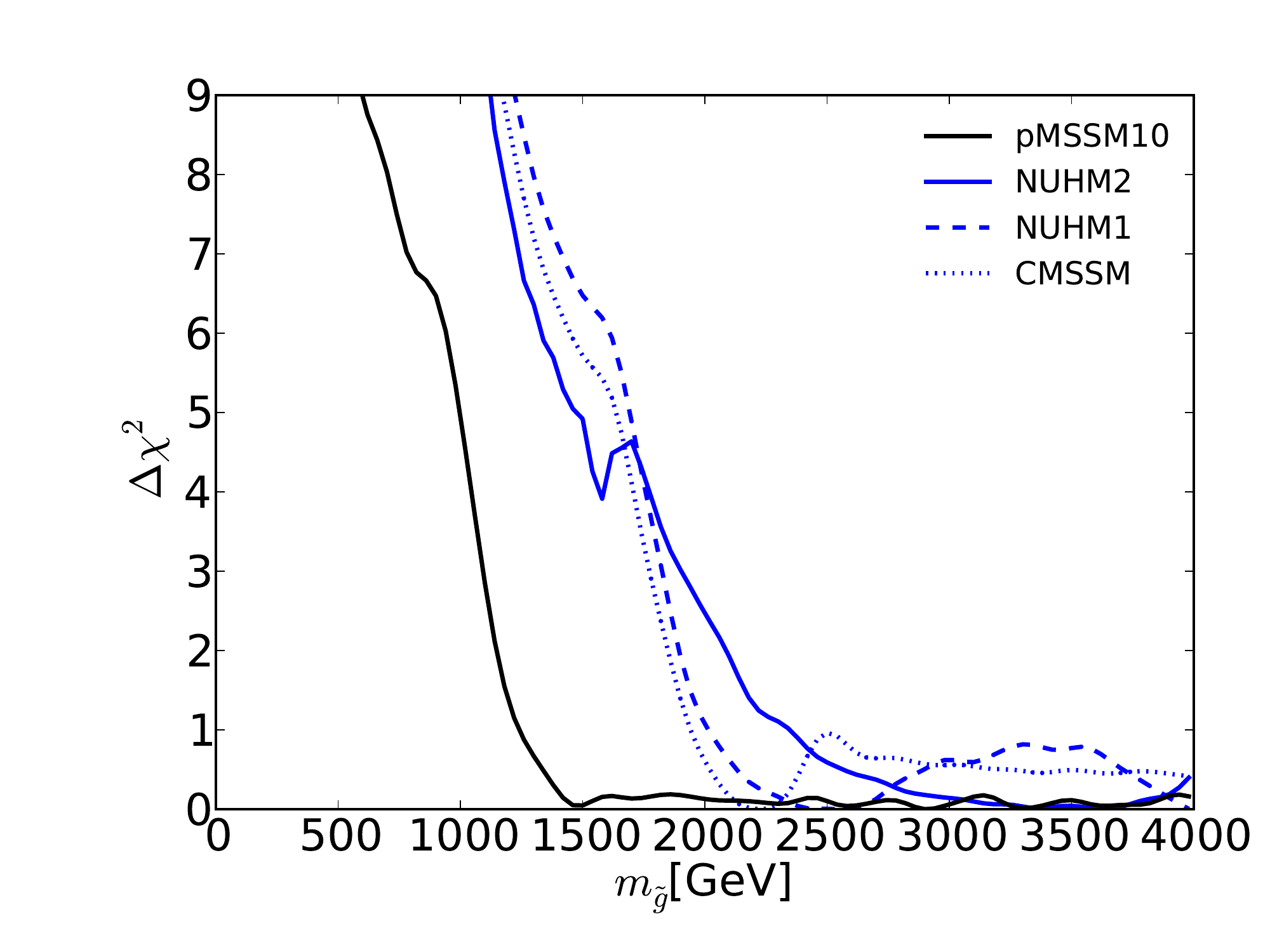}
\includegraphics[height=4cm]{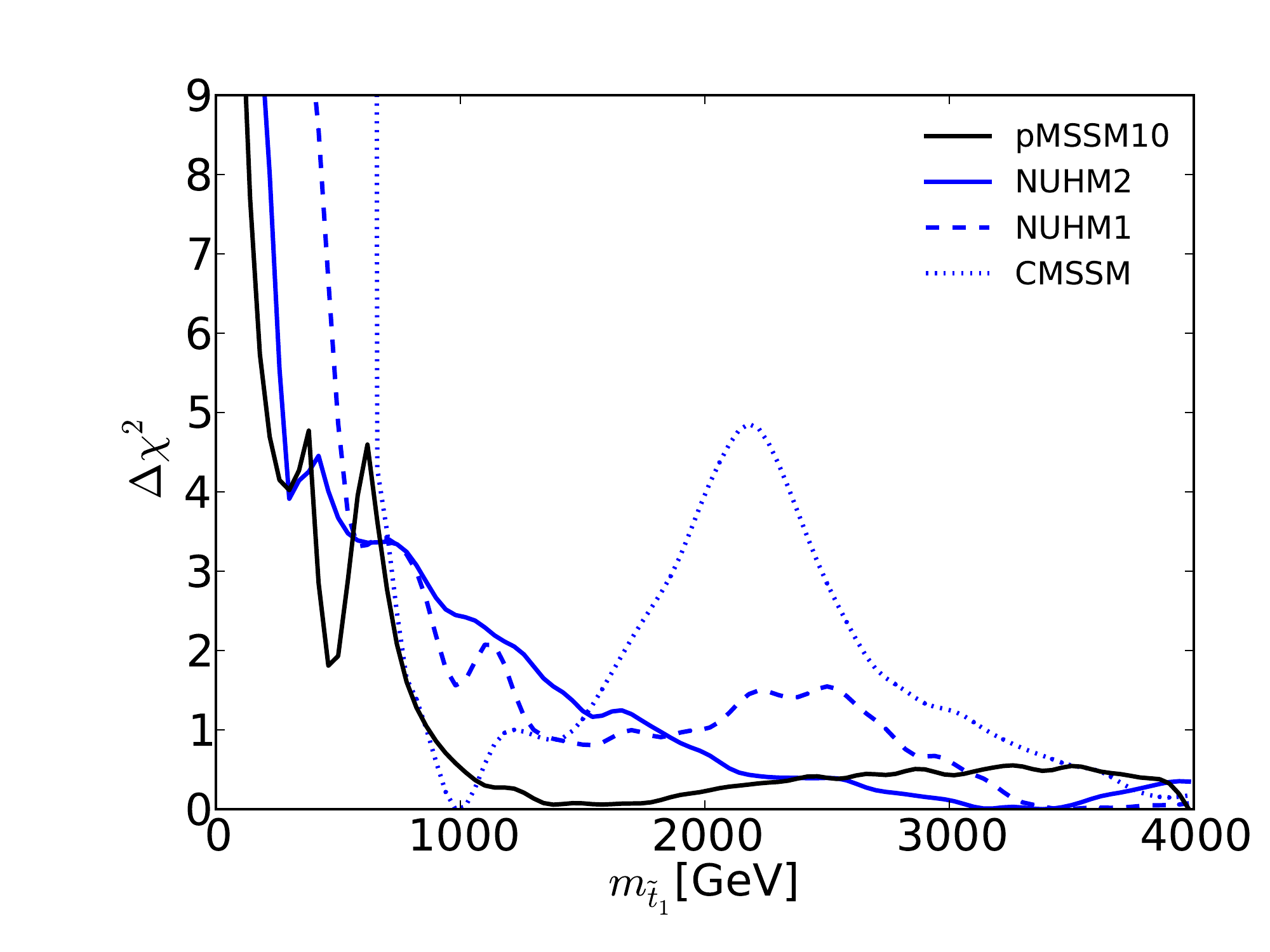} \\
\end{center}   
\caption{\label{fig:smasses}\it 
The $\chi^2$ likelihood functions for the gluino mass (left panel) and the lighter stop squark (right panel)
in the CMSSM, NUHM1, NUHM2 and pMSSM10, taking account of LHC Run~1 and other constraints,
as described in~\protect\cite{MCpMSSM}.
}
\end{figure}

The left panel of Fig.~\ref{fig:smasses} displays the dependences on the gluino mass, $m_{\tilde g}$,
of the $\chi^2$ functions from global fits to the CMSSM and related models (blue curves) and the pMSSM10
(black curve). We see that the LHC data, in particular, set 95\% CL constraints $m_{\tilde g} \gtrsim 1.5$~TeV
in the CMSSM and related models, which may be relaxed to $m_{\tilde g} \gtrsim 1.0$~TeV in the pMSSM10.
The good news is that future runs of the LHC should have sensitivity to $m_{\tilde g} \lesssim 3$~TeV, so
there are significant chances that the LHC may discover supersymmetry within these scenarios, though no
guarantees. The right panel of Fig.~\ref{fig:smasses} displays the the dependences of the of the global $\chi^2$ functions
 on the lighter stop squark mass, $m_{\tilde t_1}$, in the same line styles as in the left panel. In this case, we see
 that a `natural' light stop with $m_{\tilde t_1} \sim 400$~GeV is allowed in the pMSSM10 at the $\Delta \chi^2 \simeq 2$
 level. This region may be accessible to future LHC searches for compressed sparticle spectra.
 
 Table~\ref{tab:prospects} summarises the prospects for discovering supersymmetry in the CMSSM, NUHM1,2 and pMSSM10
 either at the LHC and/or in direct dark matter search experiments, organized according to the dominant mechanism
 for bring in the dark matter density into the range allowed by cosmology~\cite{MCDM}. A hyphen (-) indicates that the corresponding
 mechanism is not important in the given supersymmetric model. We are encouraged to see that in every box
 without a hyphen there are prospects for discovering supersymmetry at the LHC and/or in a planned
 direct dark matter search experiment. No wonder we are excited about the prospects for Run~2 of the LHC! 
If supersymmetry does escape us at the LHC, a 100~TeV collider would have
great capabilities for discovering heavy squarks and/or gluinos~\cite{Interplay}.

\begin{table*}[htb!]
\tbl{
	Summary of the detectability of supersymmetry in the CMSSM, NUHM1, NUHM2 and pMSSM10
	models at the LHC in searches for $\ETslash$ events, long-lived charged particles (LL) and heavy $A/H$
	Higgs bosons, and in direct DM search experiments, according to the dominant mechanism for
	bringing the DM density into the cosmological range~\protect\cite{MCDM}. The symbols $\checkmark$, ($\checkmark$) and
	$\times$ indicate good prospects, interesting possibilities and poorer prospects, respectively.
	The symbol -- indicates that a DM mechanism is not important for the corresponding model.}	{
	{\begin{tabular}{ | c || c || c | c | c | c|}
		\hline
		DM & Exp't & \multicolumn{4}{c|}{Models} \\ 
		mechanism & & CMSSM & NUHM1 & NUHM2 & pMSSM10 \\ \hline
		${\staue}$ & LHC & {$\checkmark$ $\ETslash$, $\checkmark$ LL} & ($\checkmark$ $\ETslash$, $\checkmark$ LL) & ($\checkmark$ $\ETslash$, $\checkmark$ LL) & ($\checkmark$ $\ETslash$), $\times$ LL \\ 
		coann. & DM & ($\checkmark$) & ($\checkmark$) & $\times$ & $\times$ \\ \hline
		$\cha{1}$ & LHC & -- & $\times$ &  $\times$ & ($\checkmark$ $\ETslash$)  \\ 
		coann. & DM & -- & $\checkmark$ & $\checkmark$ & ($\checkmark$)  \\ \hline
		${\tilde t_1}$ & LHC & -- & -- & $\checkmark$ $\ETslash$ & --  \\ 
		coann. & DM & -- & -- & $\times$ & --  \\ \hline
		$A/H$ & LHC & $\checkmark$ $A/H$ & ($\checkmark$ $A/H$) & ($\checkmark$ $A/H$) & --  \\
		funnel & DM & $\checkmark$ & $\checkmark$ & ($\checkmark$) & -- \\ \hline
		Focus & LHC & ($\checkmark$ $\ETslash$) & -- & -- & --  \\
		point & DM & $\checkmark$ & -- & -- & -- \\ \hline
		$h,Z$ & LHC & -- & -- & -- & ($\checkmark$ $\ETslash$) \\
		funnels & DM & -- & -- & -- & ($\checkmark$) \\ \hline
			\end{tabular}}}
	\label{tab:prospects}
\end{table*}

\section{Who Ordered That?}

This is the famous quip by I.~I.~Rabi about the muon. The same might be said about the
$\gamma \gamma$ `bump' with an invariant mass $\simeq 750$~GeV reported by the ATLAS~\cite{ATLASX}
and CMS experiments~\cite{CMSX} in a preliminary analysis of their 13-TeV data in December 2015. 
Both experiments now also report insignificant hints in their 8-TeV data.
At the time of writing, the data shown by ATLAS
at the Moriond conference in early March 2016
exhibit a $3.9 \sigma$ enhancement~\cite{ATLASM}, whereas the CMS data display a $3.4 \sigma$
enhancement~\cite{CMSM}. A naive combination of the p-values of the two peaks corresponds to a
$4.99 \sigma$ signal, whose significance is reduced by the `look-elsewhere effect' to
$3.89 \sigma$. This is insufficient to claim a discovery, but according to CERN Director-General
Fabiola Gianotti, we ``are allowed to be slightly excited"~\cite{FG}.

If interpreted as a new particle $X$, the reported signal would correspond to 
$\sigma(pp \to X) \times {\rm BR}(X \to \gamma \gamma) \sim$~few~fb.
Needless to say, any such $X(750)$ would itself definitely constitute physics
beyond the Standard Model, though what role it may play in resolving any of the oft-touted
outstanding problems of the Standard Model is most unclear. Even more exciting than
the existence of $X(750)$ itself is the prospect that it would be merely the tip of an
iceberg of new physics, a harbinger of a whole new layer of matter~\cite{everybody}.

Let us be conservative, and assume that the $X(750)$ has spin zero~\cite{EEQSY}. In this case,
its $\gamma \gamma$ decays would presumably be mediated by anomalous triangle
diagrams of massive charged particles. Fermions may be the most plausible candidates,
as scalar loops generally have smaller numerical values, and postulating new charged
vector bosons is disfavoured by Occam's razor. The form factors for loop diagrams are suppressed
for light fermions in the loops with masses $\ll m_X/2$, such as the top quark,
and are maximized for fermions with masses $\sim m_X/2$. However, a heavy conventional
fourth generation is strongly excluded by other constraints, and would require non-perturbative
Yukawa couplings. The most likely possibility seems to be one or more vector-like
fermions, which may have masses larger than the electroweak scale. If some of 
these are coloured, they could also mediate $X$ production via gluon-gluon fusion, which
would accommodate the energy dependence of the signal more easily than light ${\bar q} q$
collisions.

The minimal model is (1) a single vector-like charge-2/3 quark (a single bottom-like
quark would make a contribution to the $\gamma \gamma$ decay rate that is smaller
by a factor 4). Alternatively, one could postulate (2) an SU(2) doublet of vector-like quarks,
or (3) a doublet and two singlet vector-like quarks. Finally, one may go the whole hog, and
postulate (4) a full vector-like generation, including leptons as well as quarks~\cite{EEQSY}.

\begin{figure}[t!]
\begin{center}
\includegraphics[height=6cm]{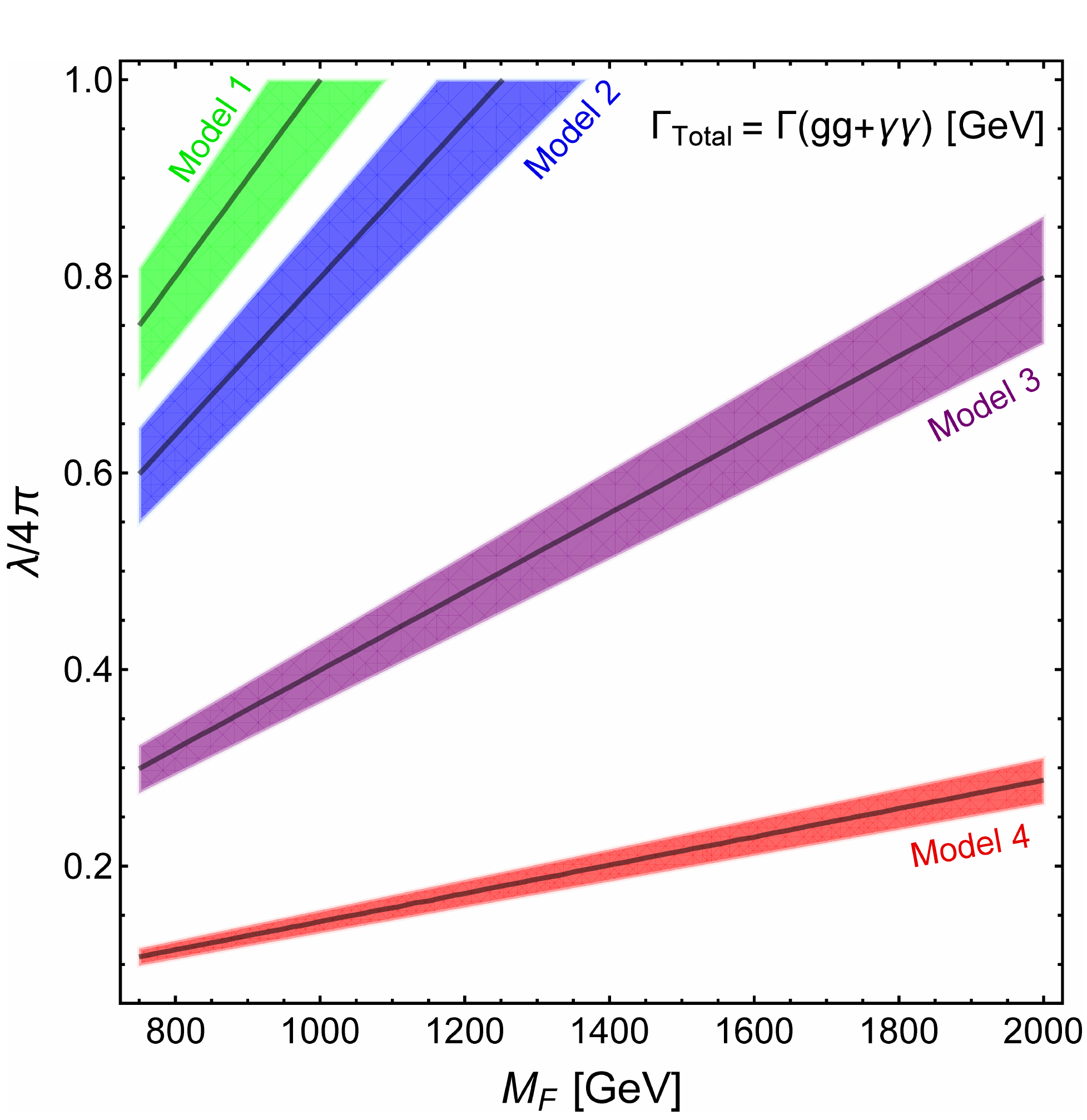}
\end{center}   
\caption{\label{fig:couplings}\it 
The $X {\bar F} F$ couplings $\lambda$ required in the vector-like fermion models (1, 2, 3) and (4) described in the text
to yield $\sigma(pp \to X) \times {\rm BR}(X \to \gamma \gamma) = 6 \pm 1$~fb (solid black lines
and coloured bands), assuming that diboson decays dominate. Plot from~\protect\cite{DEGQ}, adapted from~\protect\cite{EEQSY}.
}
\end{figure}

Fig.~\ref{fig:couplings} shows the $XF{\bar F}$ couplings $\lambda$ (assumed for
simplicity to be universal) that would be required to explain the possible $X(750)$
signal in these different models, as functions of the vector-like fermion mass
(also assumed for simplicity to be universal), under the assumption that $X \to gg$
is the dominant decay mode~\cite{EEQSY}. In each panel, the black line corresponds to 
$\sigma(pp \to X) \times {\rm BR}(X \to \gamma \gamma) = 6$~fb, and the coloured
band corresponds to $\pm 1$~fb around this central value~\cite{EEQSY}. If $\lambda/4 \pi > 1/2$
the coupling $\lambda$ is non-perturbative, whereas it is perturbative for smaller values.
We see, therefore, that models (1) and (2) may well require a non-perturbative treatment,
whereas models (3) and (4) could well be perturbative~\footnote{On the other hand, all
the models would have to be non-perturbative if $\Gamma_X \simeq 45$~GeV, as slightly
(dis)favoured by ATLAS (CMS) data.}. In the case of model (4), which
includes neutral vector-like leptons, these could constitute the dark matter if the common mass 
$\lesssim 1500$~GeV.

In each of the models studied, it is possible to calculate the ratios of the decay rates
of $X \to gg, Z \gamma, W^+ W^-, ZZ$ and $\gamma \gamma$ via the triangular loop
diagrams, with the results shown in Table~\ref{tab:BRs}. Also shown in this Table are
the upper limits on these ratios inferred from LHC 8-TeV data, as discussed in~\cite{EEQSY}.
We see that model (2) is formally in conflict with the upper limits on $X \to Z \gamma$ and $W^+ W^-$,
though it may be premature to conclude that the model is excluded. The good news is that
the models are potentially accessible to experimental searches in other diboson channels.
As discussed below, there are also interesting possibilities to look for heavy fermions at the LHC
and future colliders~\cite{DEGQ}.
All in all, there is both experimental and theoretical work for a generation if the $X$ particle exists,
and we should know the answer to this question in 2016.

\begin{table}
\tbl{Ratios of $X$ decay rates for the various models introduced in the text,
assuming $\alpha_s (m_X) \simeq 0.092$. The upper limits on $\frac{\text{BR}(X\to VV)}{\text{BR}(X\to \gamma \gamma)}$
are obtained from LHC 8-TeV data, as described in~\protect\cite{EEQSY}.}
{\begin{tabular}{c c c c c}
\textbf{Model} & $\frac{\text{BR}(X\to gg)}{\text{BR}(X\to \gamma \gamma)}$ & $\frac{\text{BR}(X\to Z \gamma)}{\text{BR}(X\to \gamma \gamma)}$ & $\frac{\text{BR}(X\to Z Z)}{\text{BR}(X\to \gamma \gamma)}$ & $\frac{\text{BR}(X\to W^\pm W^\mp)}{\text{BR}(X\to \gamma \gamma)}$ \\
\hline
\textbf{1} & 180 & 1.2 & 0.090 & 0\\
\textbf{2} & 460 & 10 & 9.1 & 61\\
\textbf{3} & 460 & 1.1 & 2.8 & 15\\
\textbf{4} & 180 & 0.46 & 2.1 & 11\\
\hline
\textbf{Current limit} & $\sim 2 \times 10^4$ &7 & 13 & 30
\end{tabular}
\label{tab:BRs}}
\end{table}

\begin{figure}[t!]
\begin{center}
\hspace{-2cm}
\includegraphics[height=4cm]{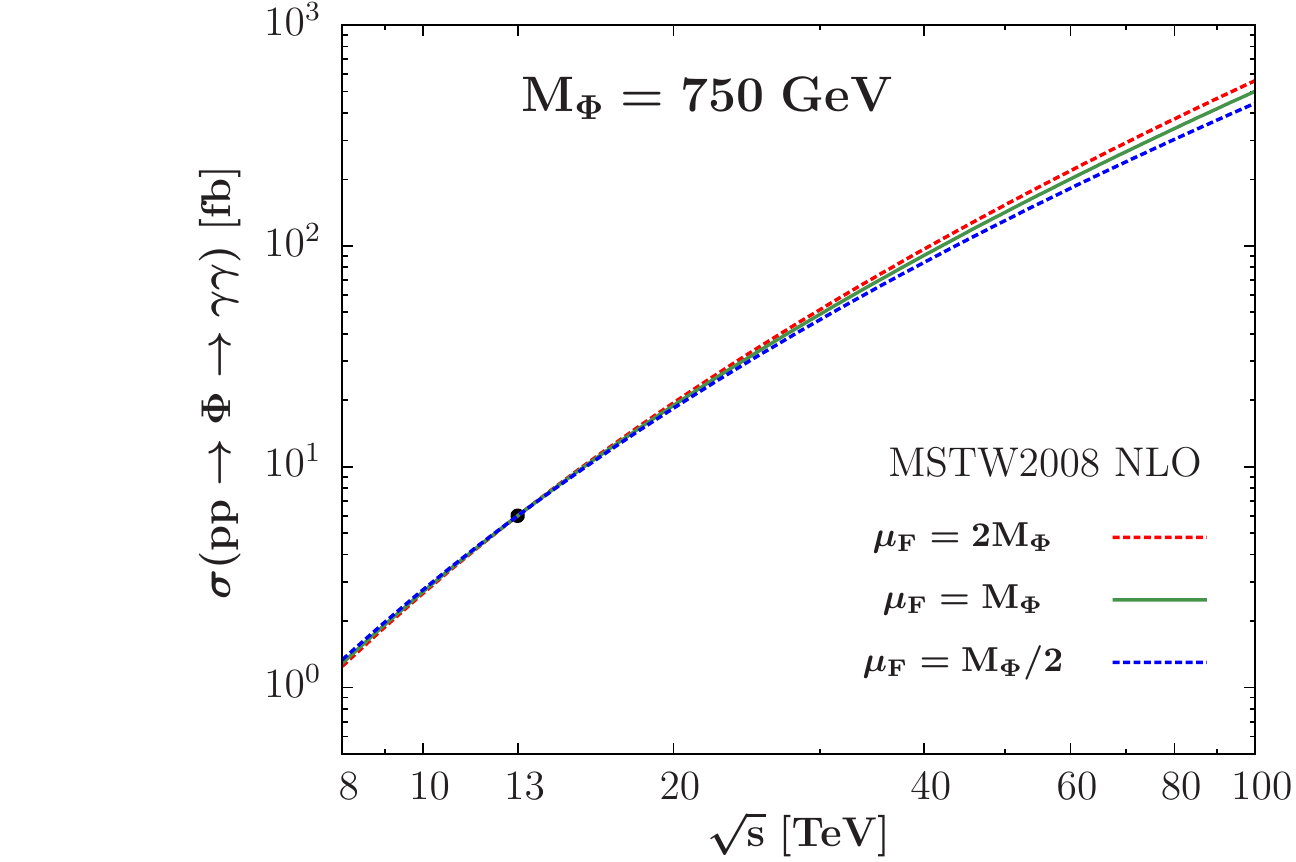}
\includegraphics[height=4cm]{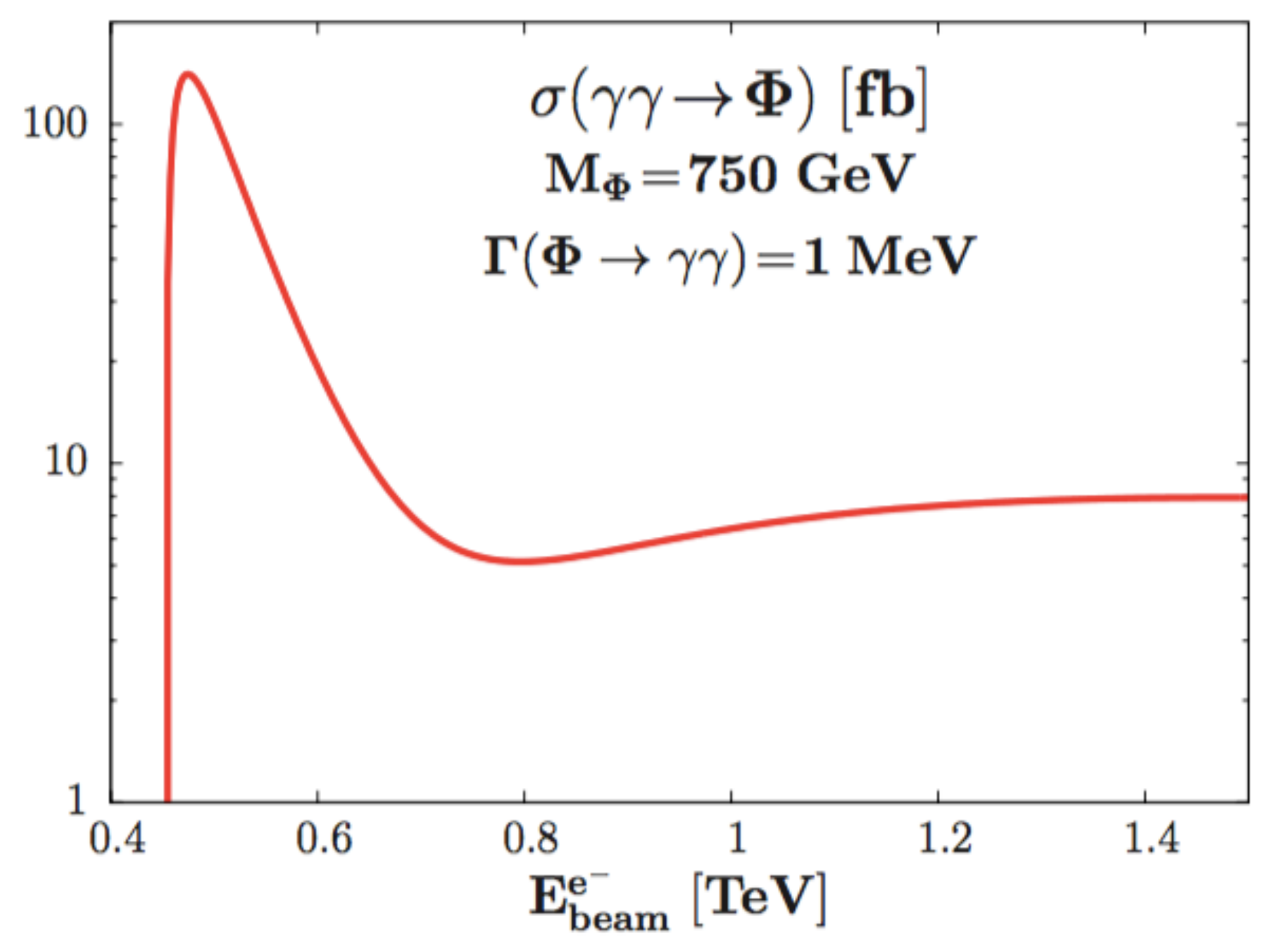} \\
\end{center}   
\caption{\label{fig:growth}\it 
Left panel: Increase of the production in $pp$ collisions at different
centre-of-mass energies of a singlet boson $\Phi$ with mass $750$~GeV produced by gluon-gluon 
collisions and decaying into $\gamma \gamma$, assuming that two-boson decays are dominant
and normalised to the possible LHC signal at $13$~TeV.
Right panel: Cross-section for its production in $\gamma \gamma$ collisions at an $e^+ e^-$
collider as a function of the electron beam energy. Plots from~\protect\cite{DEGQ}.
}
\end{figure}

The left panel of Fig.~\ref{fig:growth} shows how rapidly $\sigma(pp \to X) \times {\rm BR}(X \to \gamma \gamma)$
would grow with the $pp$ centre-of-mass energy, assuming production via gluon-gluon fusion~\cite{DEGQ}.
At 100~TeV the cross-section would increase by two orders of magnitude, with PDF and higher-order 
QCD uncertainties that are $\sim 30$\%. The right panel of Fig.~\ref{fig:growth} displays, as a function
of the $e^-$ beam energy, the cross-section for $\gamma \gamma \to X$ production at an $e^+ e^-$ 
collider that is optimised for $\gamma \gamma$ collisions. Needless to say, an $e^+ e^-$ collider 
with $E^{e^-}_{\rm Beam} < 375$~GeV would not be able to produce the $X(750)$, and we see that 
$E^{e^-}_{\rm beam} \simeq 500$~GeV would be preferred. 

\begin{figure}[t!]
\begin{center}
\hspace{-1cm}
\includegraphics[height=12cm]{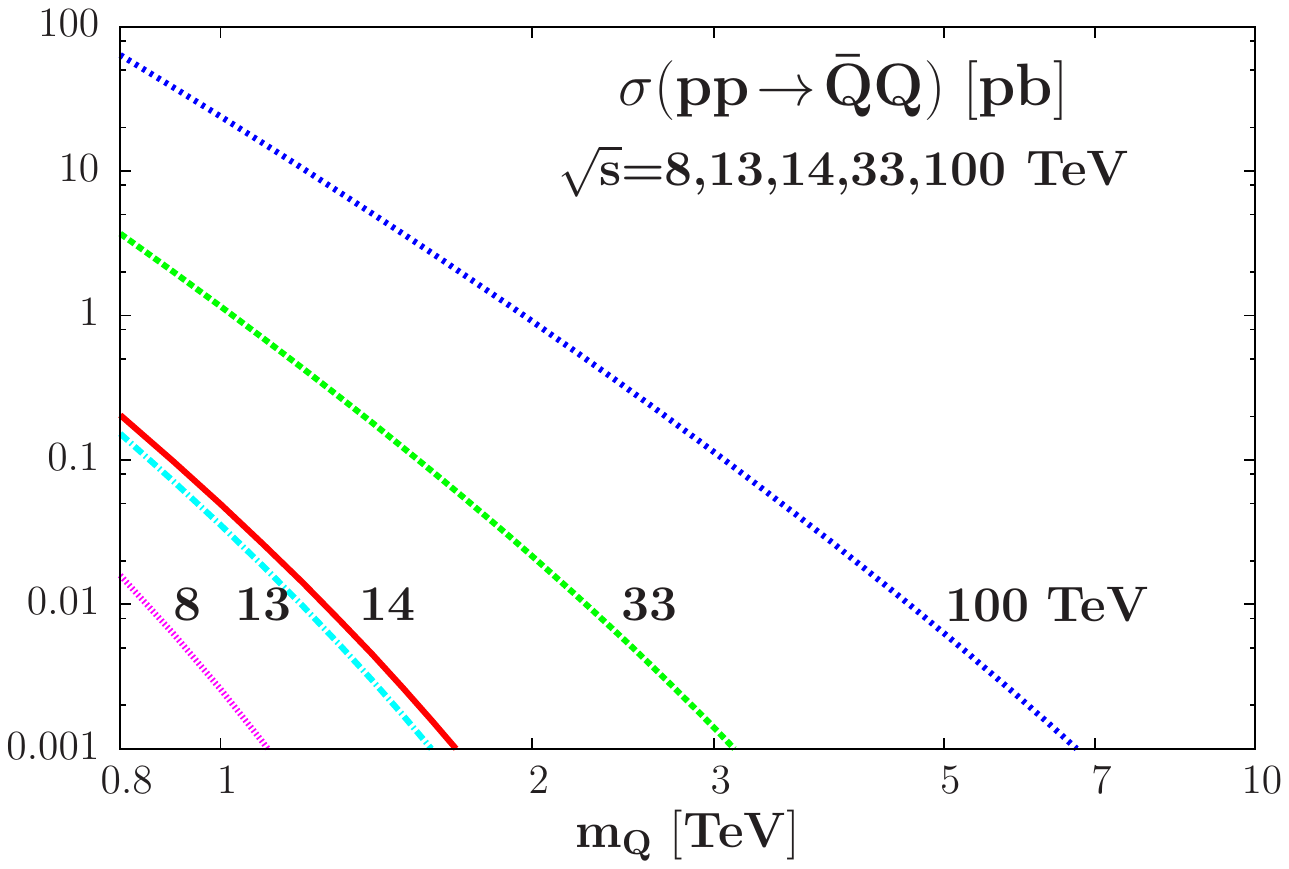}
\hspace{-4cm}
\includegraphics[height=12cm]{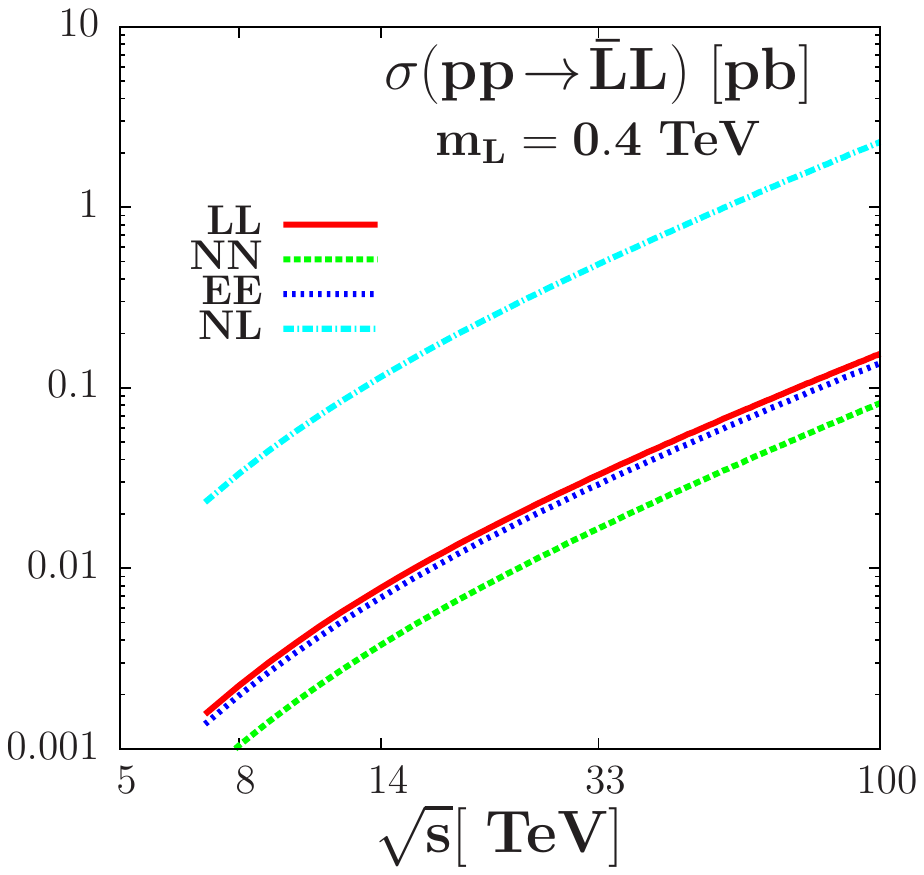} \\
\end{center}   
\vspace{-7cm}
\caption{\label{fig:VLF}\it 
Left panel: Cross-sections for vector-like quark pair-production in $pp$ collisions at different
centre-of-mass energies. Right panel: Cross-sections for the pair-production of vector-like leptons with
masses $0.4$~TeV in $pp$ collisions as functions of the centre-of-mass energy. Plots from~\protect\cite{DEGQ}.
}
\end{figure}

Fig.~\ref{fig:VLF} displays the cross-sections for the production of vector-like fermions
in $pp$ collisions as functions of the centre-of-mass energy~\cite{DEGQ}. The left panel
shows the cross-sections for vector-like quark production at different collider centre-of-mass energies
as functions of the quark mass, and the right panel shows the cross-sections for producing different
types of vector-like leptons (doublets $L$, charged and neutral leptons $E, N$ and associated $N, L$ pairs) as functions
of the centre-of-mass energy for a mass of 0.4~TeV. As we see in Table~\ref{tab:heavyF}, the LHC sensitivity for
vector-like quarks in the models (1) to (4) introduced previously
should extend to $\sim 2$~TeV and for vector-like leptons in model (4) to 0.7~TeV, and the corresponding
sensitivities of a 100~TeV collider would be to $\sim 13$ and $\sim 5$~TeV, respectively. The LHC should be able
to explore the possible range of vector-like quark masses in plausible models of $X(750)$ production and
decay, and a 100-TeV collider would be able to explore their dynamics in some detail, e.g., probing how they
mix with the Standard Model quarks~\cite{DEGQ}.

\begin{table}[!h] 
\tbl{Prospective sensitivities to vector-like quarks (left) and leptons (right) [particle masses indicated in TeV] for
various $pp$ collider scenarios.}
{\begin{tabular}{| c | c | c |}
\hline
& Vector-like quark mass sensitivity & Vector-like lepton mass sensitivity\\
{model} & 100fb$^{-1}$ ~300fb$^{-1}$ ~300fb$^{-1}$ ~20ab$^{-1}$ & 100fb$^{-1}$ 
~300fb$^{-1}$ ~300fb$^{-1} ~$20ab$^{-1}$\\
& 13~TeV ~~14~TeV ~~33 TeV ~100~TeV & 13~TeV ~~14~TeV ~~33$\;$TeV ~100$\;$TeV\\
\hline
\textbf{1} \! &\! ~~~~1.4~~~~  ~~~~1.7~~~~ ~~~~3.1~~~~ ~~~11.7~~~ & -\\
\hline
\textbf{2} \! &\! ~~~~1.5~~~~  ~~~~1.8~~~~ ~~~~3.4~~~~ ~~~12.7~~~ & -\\
\hline
\textbf{3} \! &\! ~~~~1.6~~~~  ~~~~2.0~~~~ ~~~~3.7~~~~ ~~~13.7~~~ & -\\
\hline
\textbf{4} \! &\! ~~~~1.6~~~~  ~~~~2.0~~~~ ~~~~3.7~~~~ ~~~13.7~~~ & \hspace*{-3mm}
             0.56~~  ~~0.73~~~~ ~~~1.7~~~ ~~5.3\\
\hline
\end{tabular}}
\label{tab:heavyF}
\vspace*{-4mm}
\end{table}

\begin{figure}[t!]
\begin{center}
\hspace{-1cm}
\includegraphics[height=12cm]{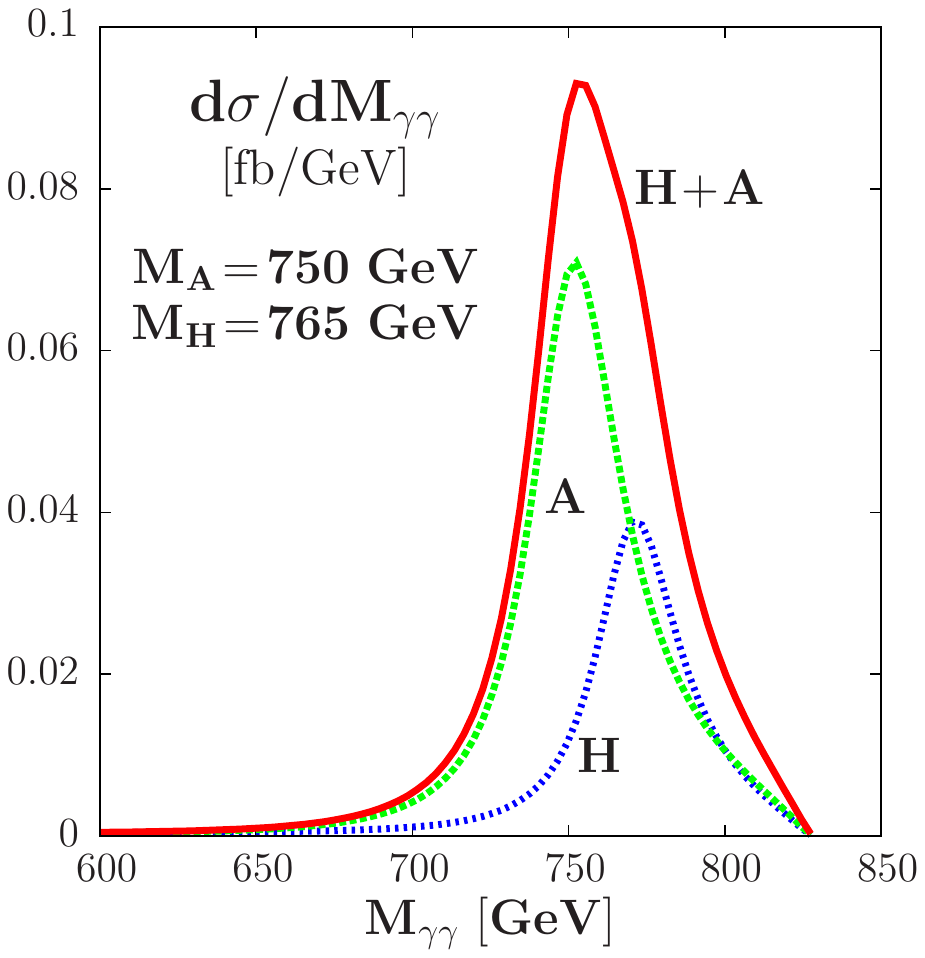}
\hspace{-4cm}
\includegraphics[height=12cm]{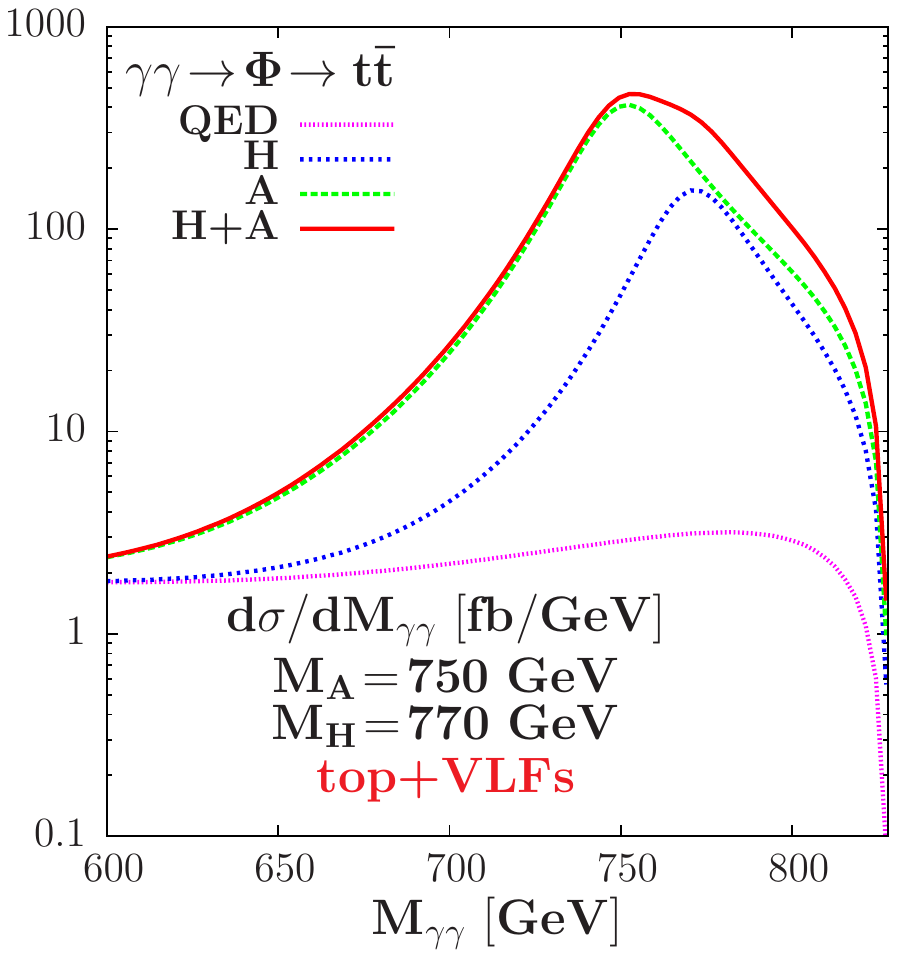} \\
\end{center}   
\vspace{-7cm}
\caption{\label{fig:lineshape}\it 
Left panel: The $\Phi = H, A$ line-shape for $M_A = 750$~GeV, $M_H= 765$~GeV,  $\Gamma_H = 32$~GeV and $\Gamma_A = 35$~GeV for $\tan \beta = 1$. Right panel: The invariant mass distribution ${\rm d\sigma/d}M_{\gamma \gamma}$
for the process $\gamma\gamma \to t\bar t$ in the $\gamma\gamma$ mode of a
linear $e^+e^-$ collider, with $H/A$ parameters as in the left panel. Plots from~\protect\cite{DEGQ}.
}
\end{figure}

As an alternative to the minimal singlet scenario for the $X(750)$ enhancement.
one may also consider a two-Higgs-doublet scenario~\cite{DEGQ}, in which it could be interpreted
as a superposed pair of heavy Higgs bosons $H, A$. In many such models, such
as supersymmetry, these bosons are nearly degenerate. For example, if the Higgs
potential is the same as in the minimal supersymmetric extension of the
Standard Model, one finds $M_H - M_A \simeq 15$~GeV if $\tan \beta = 1$.
This choice is motivated by consideration of the dominant $H/A \to {\bar t} t$ decays,
which yield $\Gamma_{H, A} = 32, 35$~GeV in this case~\footnote{Since $H/A \to {\bar t} t$ decays dominate over the
decays into boson pairs considered in the previous singlet scenario, the loop diagrams
responsible for $H/A$ production and decay must be enhanced compared to that scenario, e.g., by postulating
relatively light charged leptons with masses $\sim m_{H/A}/2$, or additional 
(multiply-?)charged particles.}. In this model, one
can also calculate the ratio $\sigma (pp \to A) \times {\rm BR}(A \to \gamma \gamma)/
\sigma (pp \to H) \times {\rm BR}(H \to \gamma \gamma) \simeq 2$. The combined
$H/A$ signal in $pp$ collisions is therefore an `asymmetric Breit-Wigner' as shown in the left panel of Fig.~\ref{fig:lineshape},
with a full width at half maximum of $\sim 45$~GeV, corresponding to the signal width
favoured (slightly) by ATLAS. The right panel of Fig.~\ref{fig:lineshape} shows the
corresponding line shape in $\gamma \gamma$ collisions at an $e^+ e^-$
collider with centre-of-mass energy 1~TeV. In addition to single $H/A$
production, there is rich bosonic phenomenology in associated ${\bar t} t H/A$
production and pair production, as discussed in~\cite{DEGQ}.

Before we get too excited, though, we should remember the wise words of Laplace: 
{\it ``Plus un fait est extraordinaire, plus il a besoin d'\^etre appuy\'e de fortes preuves"}, i.e.,
{\it ``The more extraordinary a claim, the stronger the proof required to support it"}.
The Higgs boson was (to some extent) expected, and the possible range of its mass
was quite restricted before its discovery. In contrast, the $X(750)$ is totally
unexpected. For this reason, we certainly should wait and see how the hint develops
with increased luminosity before getting much more than ``slightly excited"~\cite{FG}.

\section{Search for Sphalerons}

Let me now turn to a topic even more speculative than the existence of the $X(750)$,
namely the search for sphalerons~\cite{sphalerons}. These are non-perturbative configurations in the
electroweak sector of the Standard Model that would mediate processes that change
the SU(2) Chern-Simons number: $\Delta n \ne 0$, and thereby violate
baryon and lepton numbers, with significance for generating the cosmological
baryon asymmetry~\cite{FY}. It used to be thought that sphaleron-induced transitions would be
very suppressed at accessible energies, but this conventional wisdom has recently
been challenged by Tye and Wong (TW)~\cite{TW}. They argue that, since the effective
Chern-Simons potential is periodic, one should use Bloch wave functions $\Psi(Q)$ to 
calculate the transition rate:
\begin{equation}
\left( - \frac{1}{2 m} \frac{\partial^2}{\partial Q^2} + V(Q) \right) \Psi(Q) \; = \; E \Psi(Q) \, ,
\label{schrod}
\end{equation}
\begin{equation}
V(Q) \; \simeq \; 4.75 \left(1.31 \sin^2 (Q m_W) + 0.60 \sin^4 (Q m_W) \right) \, {\rm TeV} \, .
\label{VQ}
\end{equation}
where $Q$ is related to the Chern-Simons number by $Q \equiv \mu/m_W: n \pi = \mu - \sin(2\mu)/2$.
The Bloch wave function approach of TW yields a rate similar a tunnelling
calculation for transitions at quark-quark collision energies $E$ below the sphaleron energy 
$E_{\rm Sph} \simeq 9$~TeV, and an enhanced rate at higher energies that we parametrise as~\cite{ES}:
\begin{equation}
\sigma(\Delta n = \pm 1)  =
\frac{1}{m_W^2} \sum_{ab} \int d E  \frac{d {\cal L}_{ab}}{d E} ~p~
\exp \Big( c \frac{4 \pi}{\alpha_W} S( E ) \Big) \,,
\label{sigma}
\end{equation}
where $p$ is an unknown factor, $S(E) = 0$ for $E > E_{\rm Sph}$, and the results are
largely independent of $c$ over a plausible range.

\begin{figure}[t!]
\begin{center}
\includegraphics[height=4.5cm]{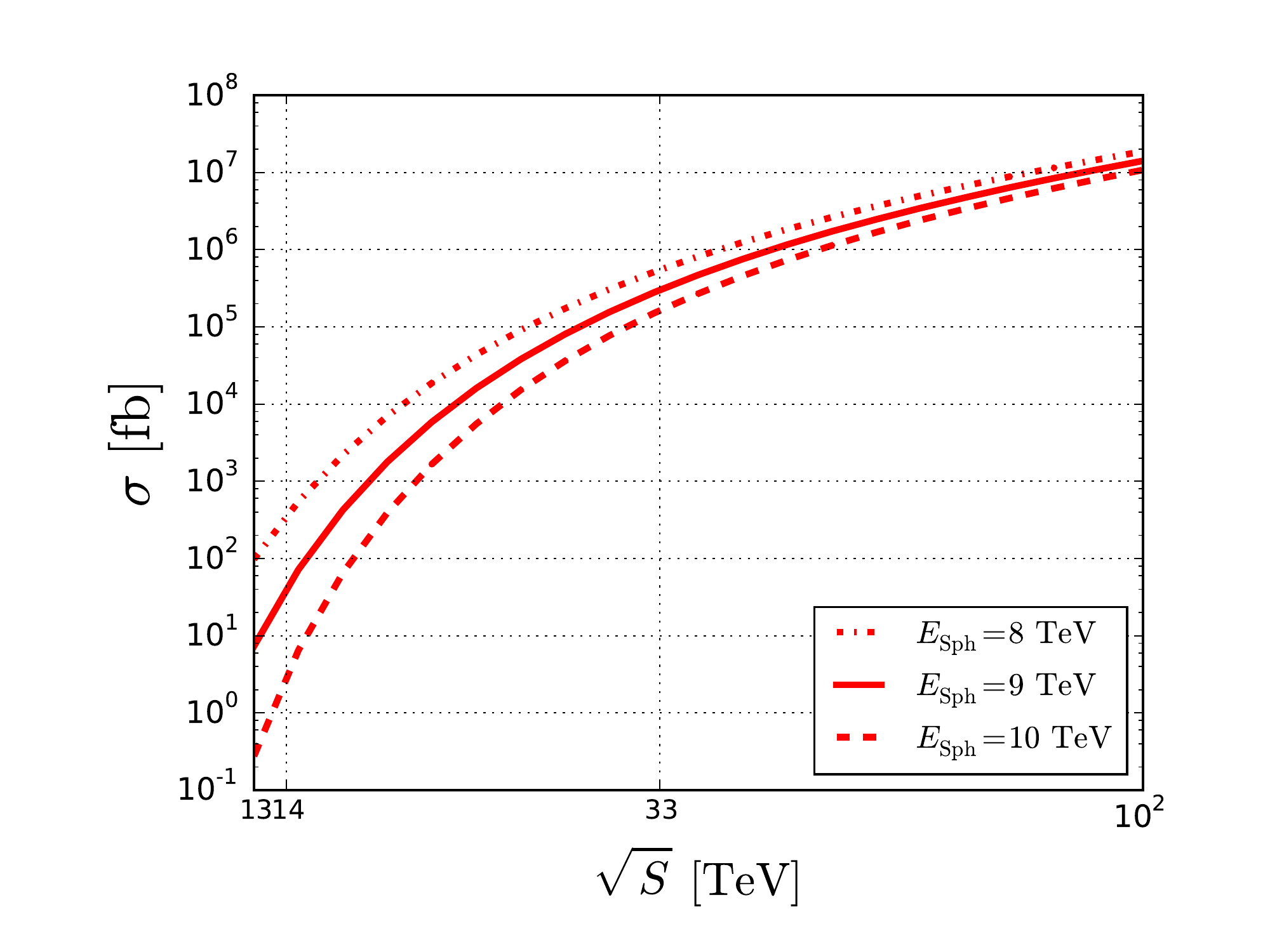}
\includegraphics[height=4.5cm]{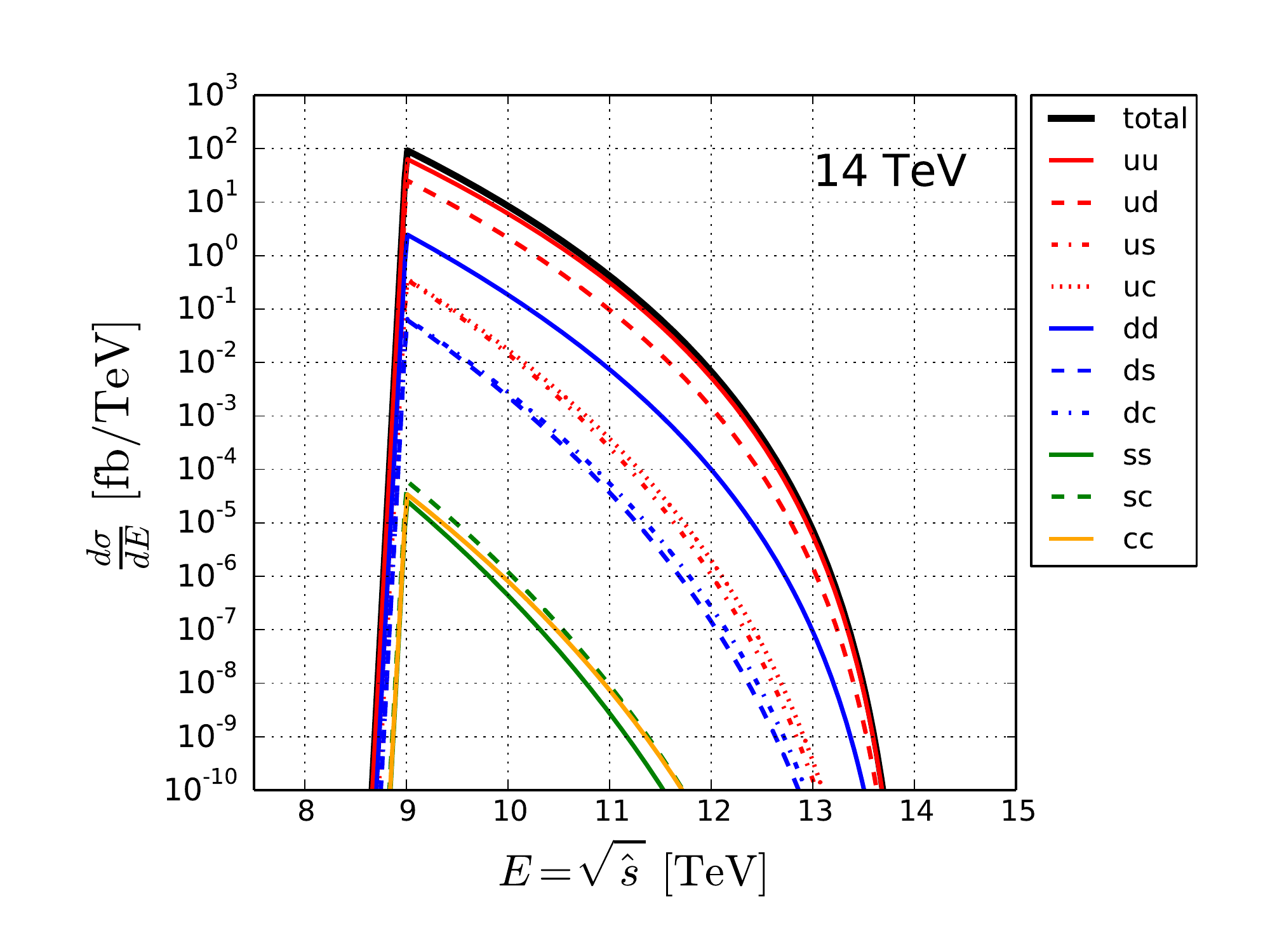} \\
\end{center}   
\vspace{-0.5cm}
\caption{\label{fig:sphaleron}\it 
Left panel: The energy dependence of the total cross section for sphaleron transitions for the nominal
choices $E_{\rm Sph} = 9$~TeV, $c = 2$ and $p = 1$ in (\protect\ref{sigma})
(solid curve), and for $E_{\rm Sph} = 8$ and $10$~TeV (dot-dashed and dashed lines, respectively).
Right panel: Contributions to the cross section for sphaleron transitions from the collisions of different flavours of quarks, 
for $E_{\rm CM} = 14$~TeV, $E_{\rm Sph} = 9$~TeV and $p = 1$ in (\protect\ref{sigma}). Plots from~\protect\cite{ES}.
}
\end{figure}

The left panel of Fig.~\ref{fig:sphaleron} shows how the sphaleron transition rate would
grow, according to (\ref{sigma}), for $E_{\rm Sph} = 9 \pm 1$~TeV~\cite{ES}. We see that the
cross-section grows significantly at the LHC between 13 and 14~TeV, and by a factor
$\sim 10^6$ between 13 and 100~TeV. It should be remembered that the normalization
factor $p$ is unknown, and that it might depend on the transition energy $E$. However,
as seen in the right panel of Fig.~\ref{fig:sphaleron}, most of the transitions take place for
$E \sim E_{\rm Sph}$, so this energy dependence may not be important.

\begin{figure}[t!]
\begin{center}
\includegraphics[height=4.6cm]{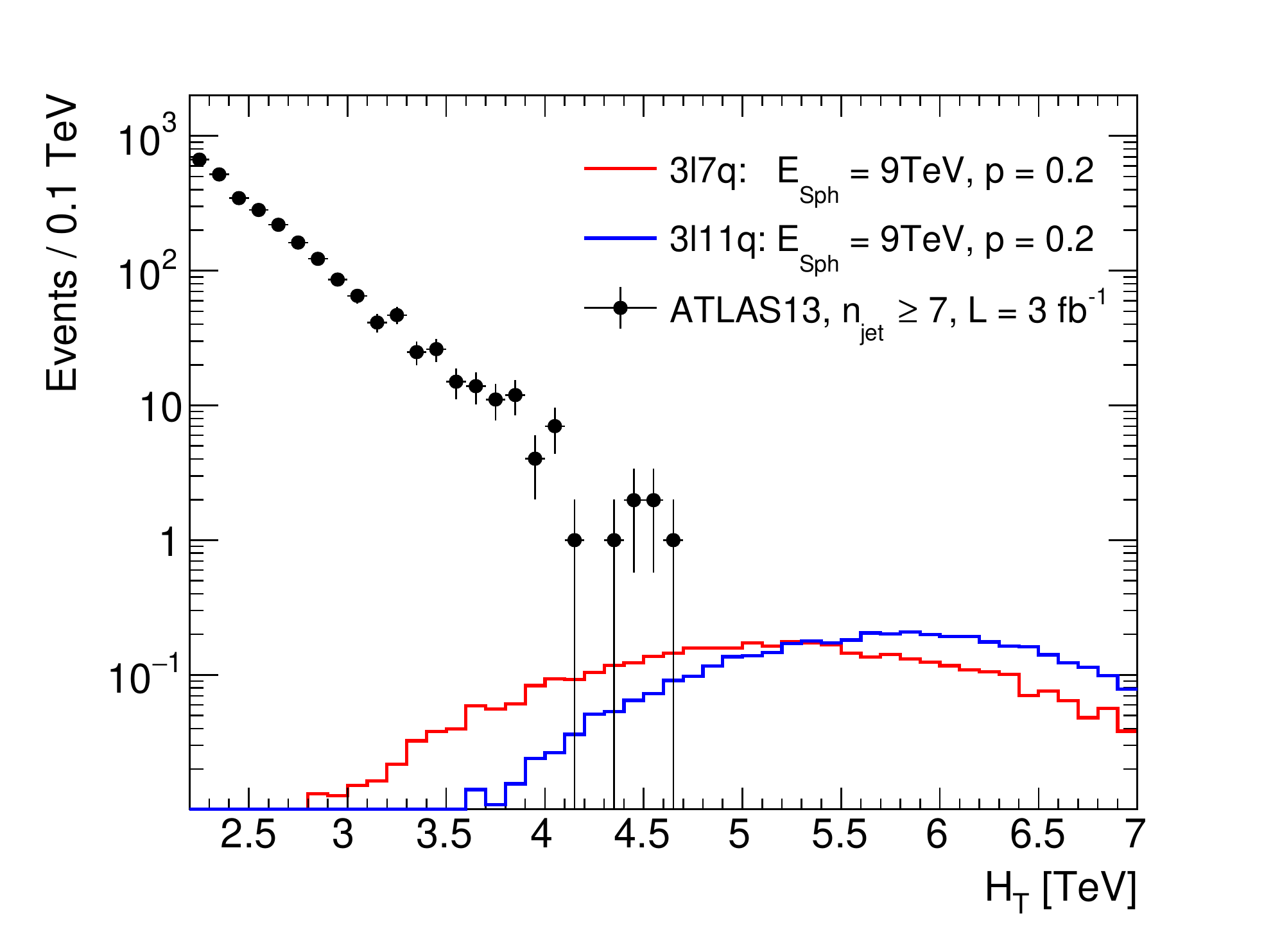}
\includegraphics[height=4.4cm]{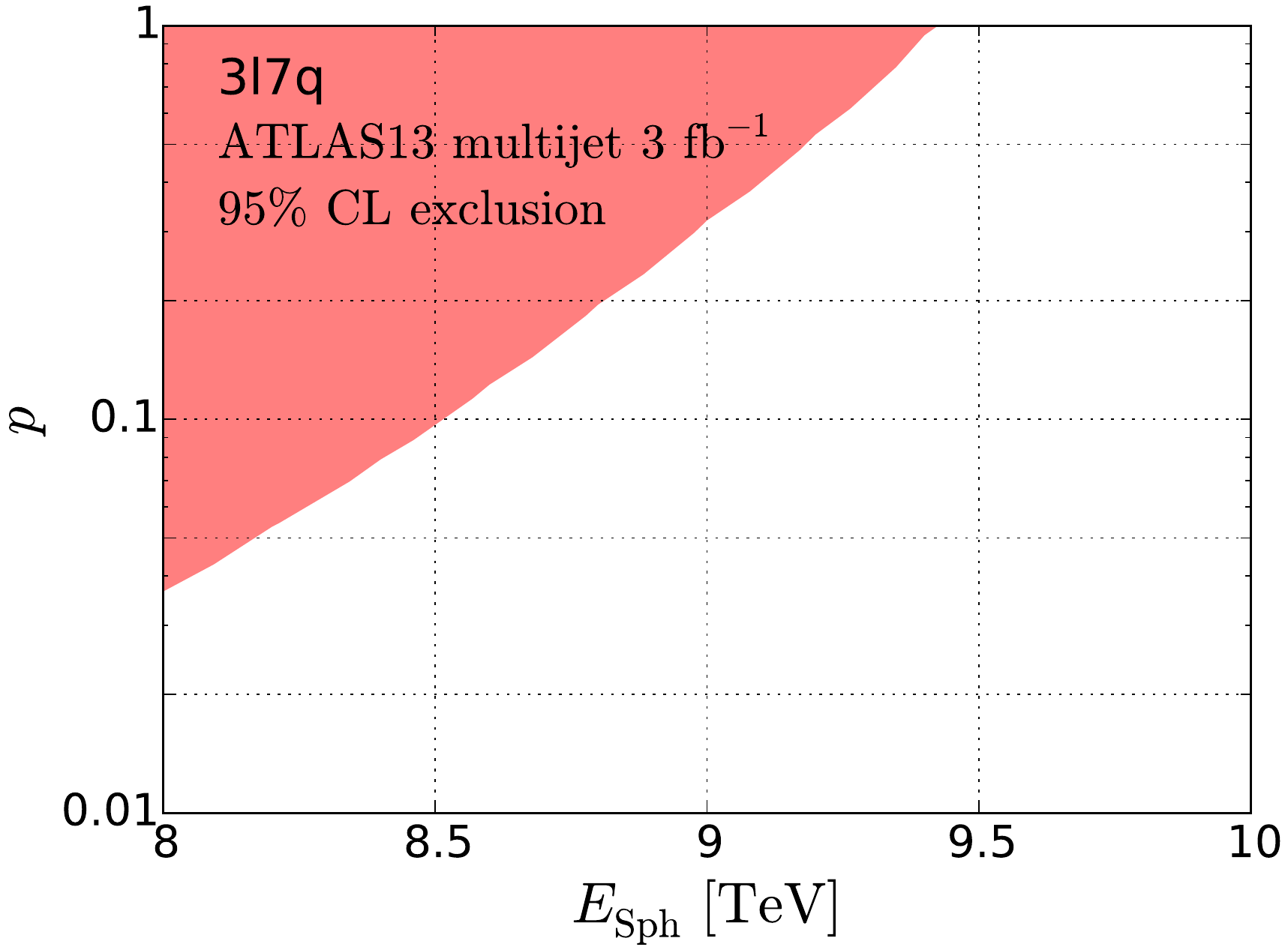}
\end{center}   
\vspace{-0.5cm}
\caption{\label{fig:BHrecast}\it 
Left panel: Comparison of the numbers of events with $n_{\rm jet} \ge 3$ measured by ATLAS in
$\sim 3$/fb of data at 13~TeV in bins of $H_T$, compared with simulations for
$E_{\rm Sph} = 9$~TeV of $\Delta n = -1$ sphaleron transitions to final states
with 3 antileptons and 7 antiquarks (red histogram) and $\Delta n = +1$ transitions to final
states with 3 leptons and 11 quarks (blue histogram). Right panel: The exclusion in the $(E_{\rm Sph}, p)$ plane for $\Delta n = - 1$ transitions obtained by recasting the ATLAS 2015 search
for microscopic black holes using $\sim 3$/fb of data at 13~TeV. Plots from~\protect\cite{ES}.
}
\end{figure}

We have simulated the final states in sphaleron-induced transitions, and found that they
are quite similar to the simulated final states for microscopic black hole decay.
Accordingly, we have recast an ATLAS search for microscopic black holes in 13-TeV
collisions with 3/fb of luminosity~\cite{ABH}, and used it to constrain the normalization factor $p$. The left panel of
Fig.~\ref{fig:BHrecast} compares the $H_T$ distribution in the final states of sphaleron
transitions with $\Delta n = \mp 1$ (labelled 3l7q and 3l11q, respectively) with the results of
the ATLAS black hole search. We see that there are no events at large $H_T$ where
the sphaleron signal would peak, and set the upper limit on $p$ shown in the right
panel of Fig.~\ref{fig:BHrecast}. The ATLAS data already set the upper limit $p \lesssim 0.3$
for $\Delta n = - 1$ transitions and the stronger constraint $p \lesssim 0.2$ for $\Delta n = + 1$ transitions
if $E_{\rm Sph} = 9$~TeV.
With 3000/fb of data at 14~TeV, the LHC would be sensitive to $p \sim 10^{-4}$, and
a 100-TeV collider with 20/ab would be sensitive to $p \sim 10^{-11}$ for $E_{\rm Sph} = 9$~TeV.
The suggestion of TW~\cite{TW} certainly needs close scrutiny, and the outcome
could open exciting prospects for future $pp$ collider experiments~\cite{ESS}.

\section{Summary}

In my opinion, rumours of the death of supersymmetry are greatly exaggerated:
it is still the most interesting framework for TeV-scale physics, and
still provides the best candidate for cold dark matter.
As discussed in this talk, simple models with universal soft supersymmetry breaking
such as the CMSSM are under pressure, with $p$-values around 0.1, but this is not
enough to reject them. More general models such as the pMSSM quite healthy,
with $p$-values around 0.3, and there are good prospects for discovering sparticles
during LHC Run 2 and/or in direct dark matter detection experiments.

More speculatively, particle physics will enter a brave new world if the $X(750)$
signal is confirmed, with exciting prospects for future $pp$ collider experiments
in particular. Let us keep our fingers crossed and await the verdict of ATLAS and
CMS during 2016.

Finally, it may be time to think again about sphalerons and the possibility that they
could have detectable effects at the LHC and future colliders.

\section*{Acknowledgements}

The author's research was supported partly by the London Centre for Terauniverse Studies (LCTS), 
using funding from the European Research Council via the Advanced Investigator Grant 26732, 
and partly by the STFC Grant ST/L000326/1. He thanks Henry Tye for hospitality at the Hong Kong
UST IAS, and Kirill Prokofiev and Luis Flores Castillo for the invitation to give this talk.

\end{document}